\documentclass[draft]{IEEEtran}
\ifCLASSINFOpdf
\else
\fi
\usepackage{slashbox}
\usepackage{amssymb}
\usepackage{amsmath}
\usepackage{color, colortbl}
\usepackage{bm,cite}
\usepackage{arydshln}
\usepackage{tabto}
\usepackage{chngcntr}
\usepackage{hyperref}
\usepackage{graphicx}
\usepackage{tabularx}
\usepackage[justification=centering, font = small]{caption}
\definecolor{Gray}{gray}{0.9}
\newtheorem{Lemma}{Lemma}
\newtheorem{Theorem}{Theorem}
\newtheorem{Definition}{Definition}

\newtheorem{Example}{Example}

\newcommand{\bc}{{\mathbf{c}}}

\newcommand{\cC}{{\mathcal{C}}}

\newcommand{\cD}{{\mathcal{D}}}

\newcommand{\cE}{{\mathcal{E}}}

\newcommand{\bG}{{\mathbf{G}}}

\newcommand{\bI}{{\mathbf{I}}}
\newcommand{\cI}{{\mathcal{I}}}

\newcommand{\bM}{{\mathbf{M}}}

\newcommand{\bP}{{\mathbf{P}}}

\newcommand{\bR}{{\mathbf{R}}}

\newcommand{\bs}{{\mathbf{s}}}

\newcommand{\cS}{{\mathcal{S}}}

\newcommand{\bT}{{\mathbf{T}}}

\newcommand{\bW}{{\mathbf{W}}}

\newcommand{\bx}{{\mathbf{x}}}

\newcommand{\al}{\alpha}

\newcommand{\del}{\delta}

\newcounter{actr}
{\begin{list}{(\alph{actr})}{\usecounter{actr}}}{\end{list}}

\newcounter{ictr}
{\begin{list}{(\roman{ictr})}{\usecounter{ictr}}}{\end{list}}

 {\everymath{\scriptscriptstyle\everymath{}}\array}%
 {\endarray}

\newcolumntype{C}[1]{>{\centering\let\newline\\\arraybackslash\hspace{0pt}}m{#1}}
\newcolumntype{L}[1]{>{\raggedright\let\newline\\\arraybackslash\hspace{0pt}}m{#1}}
\newcolumntype{R}[1]{>{\raggedleft\let\newline\\\arraybackslash\hspace{0pt}}m{#1}}

\newtheorem{remark}{Remark}

\newenvironment{new-proof}[1]
{{\em Proof }:\\}%
{ \noindent\qed }
\newcommand{\mrm}{\mathrm}

\newcommand{\floor}[1]{\lfloor{#1}\rfloor}

\DeclareMathAlphabet{\mathbsf}{OT1}{cmss}{bx}{n}
\DeclareMathAlphabet{\mathssf}{OT1}{cmss}{m}{sl}

\DeclareSymbolFont{bsfletters}{OT1}{cmss}{bx}{n}
\DeclareSymbolFont{ssfletters}{OT1}{cmss}{m}{n}
\DeclareMathSymbol{\bsfGamma}{0}{bsfletters}{'000}
\DeclareMathSymbol{\ssfGamma}{0}{ssfletters}{'000}
\DeclareMathSymbol{\bsfDelta}{0}{bsfletters}{'001}
\DeclareMathSymbol{\ssfDelta}{0}{ssfletters}{'001}
\DeclareMathSymbol{\bsfTheta}{0}{bsfletters}{'002}
\DeclareMathSymbol{\ssfTheta}{0}{ssfletters}{'002}
\DeclareMathSymbol{\bsfLambda}{0}{bsfletters}{'003}
\DeclareMathSymbol{\ssfLambda}{0}{ssfletters}{'003}
\DeclareMathSymbol{\bsfXi}{0}{bsfletters}{'004}
\DeclareMathSymbol{\ssfXi}{0}{ssfletters}{'004}
\DeclareMathSymbol{\bsfPi}{0}{bsfletters}{'005}
\DeclareMathSymbol{\ssfPi}{0}{ssfletters}{'005}
\DeclareMathSymbol{\bsfSigma}{0}{bsfletters}{'006}
\DeclareMathSymbol{\ssfSigma}{0}{ssfletters}{'006}
\DeclareMathSymbol{\bsfUpsilon}{0}{bsfletters}{'007}
\DeclareMathSymbol{\ssfUpsilon}{0}{ssfletters}{'007}
\DeclareMathSymbol{\bsfPhi}{0}{bsfletters}{'010}
\DeclareMathSymbol{\ssfPhi}{0}{ssfletters}{'010}
\DeclareMathSymbol{\bsfPsi}{0}{bsfletters}{'011}
\DeclareMathSymbol{\ssfPsi}{0}{ssfletters}{'011}
\DeclareMathSymbol{\bsfOmega}{0}{bsfletters}{'012}
\DeclareMathSymbol{\ssfOmega}{0}{ssfletters}{'012}

\begin{document}

\newcommand{\F}{\mathbb F}
%
\title{An Explicit Construction of Optimal Streaming Codes for Channels with Burst and Arbitrary Erasures}
%
%
%
\onecolumn
\author{Damian~Dudzicz,~\IEEEmembership{Student Member,~IEEE,}
        Silas~L.~Fong,~\IEEEmembership{Member,~IEEE,}
        and~Ashish~Khisti,~\IEEEmembership{Member,~IEEE}
\thanks{Manuscript received March 18, 2019.}
\thanks{D. Dudzicz is with the Department of Computer Science, EPF Lausanne, CH-1015, Switzerland}
\thanks{S. L. Fong and A. Khisti are with the Department of Electrical and Compiter Engineering, University of Toronto, Toronto, ON M5S 3G4, Canada}}

\maketitle
\begin{abstract}
This paper presents a new construction of error correcting codes which achieves optimal recovery of a streaming source over a packet erasure channel. The channel model considered is the sliding window erasure model, with burst and arbitrary losses,  introduced by Badr et al. \cite{BKTA2013}. Recently, two independents works by Fong et al.~\cite{FKLTZA2018} and Krishnan and Kumar~\cite{KK2018} have identified optimal streaming codes within this framework. In this paper, we introduce  streaming code when the rate of the code is at least $1/2$. 
Our proposed construction is explicit and systematic, uses  off-the-shelf maximum distance separable (MDS) codes and maximum rank distance (MRD) Gabidulin block codes as constituent codes and achieves the optimal error correction. 
It presents a natural generalization to the construction of Martinian and Sundberg in \cite{MartinianSundberg2004} to tolerate an arbitrary number of sparse erasures. The field size requirement which depends on the constituent MDS and MRD codes is also analyzed. 
\end{abstract}


%
\IEEEpeerreviewmaketitle

\section{Introduction}
\label{sec:1}
%
%
%
%
\IEEEPARstart{W}{ith} the development of many interactive multimedia applications such as high-quality video conferencing and virtual reality (VR), the need for reliable transmission with strict  latency constraints is a major consideration. The International Telecommunication Union recommends the end-to-end latency for such interactive applications to be less than 150 ms \cite{onewayTransTime},\cite{StockhammerHannuksela2005}. Packet loss at the physical layer often occurs in streaming application due to the unsteady nature of the link e.g. wireless or due to packet drops at transmission points in the network such as routers or switches due to high traffic. Since losses at the physical layer are inevitable, to prevent these losses two main error control schemes are currently implemented: Automatic Repeat Request (ARQ) and Forward Error Correction (FEC). The ARQ scheme presents limitations for the low-latency interactive application in the case of long-distance communication given that each retransmission of a lost packet requires an additional round-trip delay The solution studied in this paper relies on an FEC scheme which prevents the need for retransmission of dropped packets.

Motivated by these considerations, the streaming setup studied in this paper was  introduced by Martinian and Sunderberg in \cite{MartinianSundberg2004}. However this work only considered the case when the packet losses occurred in bursts, separated by sufficiently long guard in intervals. The sliding-window burst erasure model, considered in this paper, was proposed by Badr et al. in \cite{BKTA2013} along with upper and lower bounds on the capacity of the channel.
The exact capacity in this setup was determined in independent works by Fong et al.~\cite{FKLTZA2018}  and Krishnan and Kumar~\cite{KK2018}.
The proof of Fong et al. is existential in nature and involves a recursive construction for computing coefficients in the generator matrix. 
Krishnan and Kumar \cite{KK2018} present an explicit construction based on linearized polynomials and their construction is from first principles.
For other recent works on streaming codes, see e.g., \cite{LQH2013,AdlerCassuto2017,rashmi,KuijperB16,napp2016constructing,almeida2016superregular} and references therein.

This paper presents a new class of optimal rate convolutional codes when the rate of the code is at-least $R \ge 1/2$.  Our construction uses  off-the-shelf MDS and MRD block codes as constituent codes. It uses a systematic generator matrix.  It  generalizes the construction of Martinian and Sundberg  \cite{MartinianSundberg2004} which was limited to correcting burst erasures, in a natural way.  The structure of the generator matrix is fully described for any set of parameters when $R \ge \frac{1}{2}$, and the field size requirements of the construction are discussed.

The rest of the paper is organized as follows. In Section II, we define the channel model and formulate the problem. We also state the notations being used in the paper. In Section III, we state the preliminaries required to carry the proof. We remind some properties of MDS, MRD and Gabidulin codes and the standard argument of interleaving a block code into a convolutional code. In Section IV, we present the construction of the code and we prove its error correcting properties. In Section V, we discuss the field size requirements of the construction. We also compare our construction to the others currently available in the literature

\section{Problem Statement}
\label{sec:2}
\subsection{Streaming Codes}
\label{sec:enc}
We consider a sending source $\cS$ which generates at each time instant $t \in \{0,1,2,\dots\}$ a packet \linebreak $\textbf{s}[t] \triangleq [~s_0[t], s_1[t],\dots,s_{k-1}[t]~]^\text{T}$ with $s_i[t] \in \mathbb{F}_q$ for $i \in \{0,1,2,\dots,k-1\}$. Note that $\mathbb{F}_q$ is a finite field of size $q$ such that $\textbf{s}[t] \in \mathbb{F}_q^k$.
Each source packet $\textbf{s}[t]$ is encoded using a causal convolutional encoder $\cE$ such that the encoded packet $\textbf{x}[t] = [~x_0[t],x_1[t],\dots,x_{n-1}[t]~]^\text{T}=\mathcal{E} \left( \textbf{s}[0],\textbf{s}[1],\dots,\textbf{s}[t] \right)$. Where $\textbf{x}[t] \in \mathbb{F}_q^n$ and $\mathcal{E}:\mathbb{F}_q^{k \cdot (t+1)} \rightarrow \mathbb{F}_q^n$ is the encoding function.

Each encoded packet $\textbf{x}[t]$ is transmitted over a channel which introduces erasures on a packet level. The receiver receives at each time $t \in \{0,1,2, \dots \}$ the packet $\textbf{y}[t]$ such that
\begin{align}
\textbf{y}[t] = 
\begin{cases}
* &\text{if \textbf{x}[t] is erased} \\
\textbf{x}[t] &\text{otherwise}
\end{cases}
\end{align}

At the receiver the decoder $\cD$ must reconstruct perfectly the source packet $\textbf{s}[t]$ within the delay $T$ given the previously received packets $\{\textbf{y}[0],\textbf{y}[1],\dots,\textbf{y}[t+T]\}$, i.e., \linebreak $\hat{\textbf{s}}[t] = \mathcal{D}(\textbf{y}[0],\textbf{y}[1],\dots,\textbf{y}[t+T]) = \textbf{s}[t]$ with $\mathcal{D}$ being the decoding function and $\hat{\textbf{s}}[t]$ being the reconstituted source packet $\textbf{s}[t]$ by the decoder.

\subsection{Channel Model}

\begin{figure}[h]
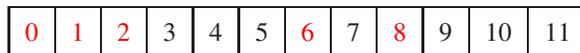

  \centering
  \captionsetup{justification=centering}
  	\begin{tabular}{|c|c|c|c|c|c|c|c|c|c|c|c|}
	\hline
 	\color{red} 0 & \color{red} 1 & \color{red} 2 & 3 & 4 & 5 & \color{red} 6 & 7 & \color{red} 8 & 9 & 10 & 11\\
	\hline
	\end{tabular}
	  \caption{A (5,3,2)-erasure channel example, with red indices indicating the corresponding erased packets}
\label{fig:1}
\end{figure}

The channel model considered in this work is the sliding-window burst errasure channel denoted by $\mathcal{C}(W,B,N)$ that was introduced by Badr et al. in \cite{BKTA2013,BPKTA17}. This model admits up to $B$ consecutive erasures or $N$ sparse non-consecutive erasures in any window of size $W$ among the sequence of transmitted packets $\textbf{x}[t]$; see Fig.~\ref{fig:1} for an example.

\indent Throughout the paper, we will assume that $W \geq T + 1$. In the setting when $B < W \le T+1$ we can achieve the capacity by reducing the effective delay to $T_\mrm{eff} = W-1$ 
as discussed in~\cite{BPKTA17}. Furthermore the capacity is trivially zero if $W \le B$ as an erasure sequence that erases all the channel packets becomes admissible.
\\
\indent Thus we can assume w.l.o.g. 
\begin{align}
W > T \geq B \geq N \label{eq:c1}
\end{align}

\subsection{Capacity}
A streaming code with the encoder and decoder definitions in Section~\ref{sec:enc} is feasible for the $\cC(N,B,W)$ sliding window channel if every source packet can be recovered with a delay of $T$. The maximum achievable rate $R = \frac{k}{n}$ of a feasible code is the capacity.
In Badr et. al~\cite{BPKTA17} upper and lower bounds on the capacity  were established:
\begin{align}
\frac{T-N}{T-N+B} \le R \le \frac{T-N+1}{T-N+B+1}. \label{eq:cap_bnds}
\end{align}
For the special case when $R=1/2$ it was already known that the upper bound is tight~\cite{BKTA2013}. However only recently independent works in~\cite{KK2018,FKLTZA2018}
established that for all parameters in~\eqref{eq:c1} the capacity is given by:
\begin{align}
\label{eq:cap}
C = \frac{T-N+1}{T-N+B+1}
\end{align}
and thus the upper bound in~\eqref{eq:cap_bnds} is indeed tight.
\section{Preliminaries}
\label{sec:3}
\subsection{Notations}

The finite field of size $q$ where the elements of the matrix live is denoted by $\mathbb{F}_q$. The extension field is denoted by $\mathbb{F}_{q^m}$. The set of all row vectors of size $k$ in the base field is denoted by $\mathbb{F}_q^k$ and the set of $k \times n$ matrices by $\mathbb{F}_q^{k \times n}$. The symbol vectors are represented using the bold character $\textbf{s}$. The generator matrix are denoted by $\textbf{G}$. The identity matrix of size $k$ is denoted by $\bI_k$. The all-zero matrix of size $A \times B$ is denoted by $\bm{0}^{A \times B}$.

In a standard manner, we define by $R = \frac{k}{n}$ the rate of the code. In this paper, we will consider high rate regime codes as codes with $R \geq \frac{1}{2}$. Using the fact that $R = \frac{k}{n}$  and $n=k+B$ (c.f.~\eqref{eq:cap}) this is equivalent to requiring that $k \geq B$. 

\subsection{Linear Block Code}

Let us denote by $\textbf{G} \in \mathbb{F}_{q^m}^{k \times n}$ the generator matrix corresponding to a given linear block code $(n,k)$, with $k$ being the dimension of the source packet $\textbf{s} \in \mathbb{F}_{q^m}^k$ and $n$ the dimension of the encoded packet $\textbf{x} \in \mathbb{F}_{q^m}^{n}$. In a linear block code, the encoding process can be expressed in terms of matrix multiplication.
\begin{align}
\textbf{x}^{\textrm{T}} &= \textbf{s}^{\textrm{T}} \cdot \textbf{G}
\\
[x_0 ~ x_1 ~ \dots ~ x_{n-1}] &= [s_0 ~ s_1 ~ \dots ~ s_{k-1}] \cdot \textbf{G}
\end{align}

Each one of such encoded symbols is successively transmitted over the erasure channel, one symbol per timeslot.

In this paper we will only consider generator matrices that are in their systematic form:
\begin{align}
\label{Def:G}
\textbf{G} = \left[ \bI_k ~~ \textbf{P} \right] 
\end{align}
with
\begin{enumerate}
\item $\bI_k$ being the identity matrix of size $k$
\item \textbf{P} being the $k \times (n-k)$ parity matrix
\end{enumerate}

In order to assert the error correction capacities of our code, we formulate the following error correction property of a linear block code.

\begin{Definition}
A $(n,k)$ linear block code is said to respect the set of constraint $(W,T,B,N)$ if there exists a set of decoding functions $\delta_{i+T}:\{\mathbb{F}_{q^m} \cup \{*\}\} ^L \rightarrow \{\mathbb{F}_{q^m} \cup \{*\}\} ^L$ with $i \in \{0,1,\dots,k-1\}$, $L = \textrm{min}\{i + T +1, n\}$ and $*$ denoting an erasure such that $s_i = \delta_{i+T}(y_0,y_1,\dots,y_{L-1})$. In other words, given $\{y_0,y_1,\dots,y_{L-1}\}$ the symbols received at the decoder up to time $t = L-1$, the block code allows perfect recovery of all the source symbols $s_i$ under a decoding delay $T$ for any pattern of $N$ sparse erasures or $B$ burst in the time window $W$. 
\end{Definition}

\subsection{MDS codes}
We remind the reader with some general results of the maximum-distance separable (MDS) codes, used in our construction.

The generator matrix of a systematic $(n,k)$-MDS code can be expressed as $\bG_\mrm{MDS} = \left[\bI_k \quad {\bf{V}}^{k \times (n-k)}\right] \in \mathbb{F}_{q}^{(k \times (n-k))}$ with ${\bf{V}}^{k \times (n-k)}$ being a Cauchy matrix. Such generator matrix is guaranteed to exist as long as $q\geq n$ and $q$ is prime \cite{MacWilliamsSloane1988}.

In our construction, we extensively use the following property of MDS codes.

\begin{Lemma}
\cite{MacWilliamsSloane1988}
A given  $(n,k)$-code $\cC$ is MDS iff any set of $k$ columns of its generator matrix $\bf{G}_{\bf{MDS}}$ are linearly independent over $\mathbb{F}_q$.
\end{Lemma}

\subsection{MRD Gabidulin codes}

Given that Gabidulin codes matrices are used in our construction, we present the reader with the main elements of the theory of maximum rank distance (MRD) and Gabidulin codes.

\begin{Definition}
A Rank Code $\cC$ is a specific matrix code which is defined as an non-empty subset of $\mathbb{F}_q^{m\times n}$ equipped with the rank distance metric $d_R(\boldsymbol{x},\boldsymbol{x}') \triangleq \textrm{Rank} (\boldsymbol{x}-\boldsymbol{x}')$ where $\boldsymbol{x},\boldsymbol{x}' \in \cC$. 
\end{Definition}

Given that $\mathbb{F}_q^{n\times m} \cong \mathbb{F}_{q^m}^{n}$, any rank code can be also represented in its corresponding block code form over the extension field $\mathbb{F}_{q^m}$.

\begin{Definition}
A Maximum Rank Distance (MRD) Code with a generator matrix $\mathbf{G} \in \mathbb{F}_{q^m}^{k \times n}$ is a code that achieves the maximum possible minimum rank distance:
 $d_{min} \triangleq \min\limits_{\boldsymbol{x},\boldsymbol{x}' \in C : \boldsymbol{x} \neq \boldsymbol{x}'} d_R(\boldsymbol{x},\boldsymbol{x}')$.
\end{Definition}

Such codes present an useful property in the scope of our construction.

\begin{Theorem}
\cite{Gab85},\cite{Neri2018} Let $\mathbf{G} \in \mathbb{F}_{q^m}^{k \times n}$ be the generator matrix of a rank code $\cC \subseteq \mathbb{F}_{q^m}^n$. Then $\cC$ is an MRD code iff for $\mathbf{T} \in \mathbb{F}_{q}^{n \times k}$ s.t. $\text{rank}\left(\mathbf{T}\right)=k$, $\mathbf{G} \cdot \mathbf{T}$ is also of rank $k$. 
\end{Theorem}

For $m \geq n$, Gabidulin introduced in \cite{Gab85} a class of such codes called Gabidulin codes. 

\begin{Definition}
	For $m \geq n$, let $g_1,g_2,\dots,g_n \in \mathbb{F}_{q^m}$ be linearly independent elements over $\mathbb{F}_{q}$. Then an $(n,k)$-Gabidulin code is defined by the following generator matrix $\bG$ 
\end{Definition}

\begin{align}
\label{matrix:gabidulin}
	\mathbf{G} = 
	\left[
		\begin{array}{c c c c}
			g_1 & g_2 & \dots & g_n
			\\
			g_1^q & g_2^q & \dots & g_n^q
			\\
			g_1^{q^2} & g_2^{q^2} & \dots & g_n^{q^2}
			\\
			\vdots & \vdots & \ddots & \vdots
			\\
			g_1^{q^{k-1}} & g_2^{q^{k-1}} & \dots & g_n^{q^{k-1}}
		\end{array}
	\right]
\end{align}

Note that Gabidulin codes are also MDS codes given that any MRD code s.t. $ n \leq m$ is also an MDS code \cite{berger2009construction}.

The following observation is immediate.
\begin{Lemma}
\label{lem:sample-gen-1}
Let $\bG \in  \mathbb{F}_{q^m}^{k\times n}$ be the generator matrix of an MRD code in~\eqref{matrix:gabidulin}. Let $\bG'\in  \mathbb{F}_{q^m}^{k\times n'}$ be a sub-matrix of $\bG$ obtained by selecting any $n' > k$ columns of $\bG$. Then $\bG'$ is the generator matrix of an $(n',k)$ MRD code.
\end{Lemma}
The proof of  Lemma~\ref{lem:sample-gen-1} follows by noting that every column in $\bG$ involves powers of a distinct element in $\mathbb{F}_{q^m}$; thus taking a subset of columns results in the same form as in~\eqref{matrix:gabidulin}.

\begin{Lemma}
\cite{Al2017} If $\cC \subseteq \mathbb{F}_{q^m}^n$ is an MRD code, then there exists for this code a generator matrix $\mathbf{G} \in \mathbb{F}_{q^m}^{k \times n}$ which can be written as
\begin{align*}
    \mathbf{G} = \left[ \bI_k ~~ \mathbf{P} \right]
\end{align*}
with $\mathbf{P}$ a parity matrix having all its entries from $\mathbb{F}_{q^m} \backslash \mathbb{F}_{q}$.
\label{lem:sys}
\end{Lemma}

Note that explicit constructions of such systematic Gabidulin generator matrix are available in the literature. In \cite{aless2018systematic}, Neri provides an explicit parametrization of systematic Gabidulin codes involving generalized rank Cauchy matrices.

We present the reader with a lemma used in the latter proof of the correctness of our construction.

\begin{Lemma}
	Any subset of $n' > k$ columns of a systematic Gabidulin code generator matrix $\mathbf{G} \in \mathbb{F}_{q^m}^{k \times n}$ yields a generator matrix $\mathbf{G}' \in \mathbb{F}_{q^m}^{k \times n'}$ of a MRD code.
	\label{lem:syscol}
\end{Lemma}

\begin{IEEEproof}
        Let $\bG$ be a $(n,k)$ Gabidulin code as in~\eqref{matrix:gabidulin} and let $\bG^\mrm{sys}$ be the generator matrix of the same code in the systematic form according to Lemma~\ref{lem:sys}. Note that since $\bG^\mrm{sys}$ can be obtained from $\bG$ by performing elementary row-operations, we can express:
        \begin{align}
        \bG^\mrm{sys} = \bM \cdot \bG  \label{eq:row-red}
        \end{align}
        where the matrix $\bM \in \mathbb{F}_{q^m}^{k \times k} $ is a full-rank matrix. 
        
        We wish to show that selecting any $n' > k$ columns of $\bG^\mrm{sys}$ also results in a MRD code.  Let $\bG'$ be the generator matrix resulting from this operation. 
        It suffices to show that for any matrix $\mathbf{T} \in \mathbb{F}_{q}^{n' \times k}$ which is full rank, $\bG' \cdot \mathbf{T}$ is also a full rank matrix.
        
        We can express:
        \begin{align}
        \bG' = \bG^\mrm{sys} \cdot \bR \label{eq:col-selec}
        \end{align}
        where  $\bR \in \mathbb{F}_{q^m}^{n \times n'} $ is a matrix whose columns consist of cartesian vectors of length $n$ associated with the selected columns.
        
        Using~\eqref{eq:row-red} and~\eqref{eq:col-selec} it suffices to show that the matrix $\mathbf{M} \cdot \mathbf{G} \cdot \mathbf{R} \cdot \mathbf{T}$ is a full rank matrix.
        Since $\bM$ is a full rank matrix this is equivalent to showing that $\mathbf{G} \cdot \mathbf{R} \cdot \mathbf{T}$ is a full rank matrix.  However note that the matrix $\mathbf{G} \cdot \mathbf{R}$ consists of selecting $n' > k$ columns of the generator matrix $\bG$ which is in the from~\eqref{matrix:gabidulin}. Using Lemma~\ref{lem:sample-gen-1} this is also a generator matrix of an MRD code. It follows that $\mathbf{G} \cdot \mathbf{R} \cdot \mathbf{T}$ is a full rank matrix whenever $\bT$ is a full rank matrix.
\end{IEEEproof}

\subsection{Diagonal Interleaving}
In order to construct the desired streaming code for the set of parameters $(W,T,B,N)$, we use the method of periodic interleaving of a linear block code into a convolutional \cite{Forney1971},\cite{MartinianSundberg2004}.

We also remind the achievability property of the periodic interleaving stated in \cite{FKLTZA2018}, without restating its proof.

\begin{Theorem}
Given an $(n,k)$ block code achieving the set of constraints $(W,T,B,N)$, it is possible to construct an $(n,k,n-1)$ convolutional code which also satisfies the same set of constrain $(W,T,B,N)$.
\end{Theorem}

To keep this paper self contained, the following is an example in Fig.~\ref{fig:exp} of the application of the periodic interleaving technique. A $(6,3)$ block code with time delay constraint $T=4$ can be diagonally interleaved in order to obtained a $(6,3,5)$ convolutional code respecting the same time delay constraint $T$. The reader can easily verify that it is indeed the case. 

\begin{figure}[!h]
\centering
\begin{tabular}{|c|*{8}{c|}}\hline
\backslashbox{\small Symbol}{\small Time}
&\makebox[1.5em]{$i-2$}&\makebox[1.5em]{$i-1$}&\makebox[1.5em]{$i$}&\makebox[1.5em]{$i+1$}&\makebox[1.5em]{$i+2$}
&\makebox[1.5em]{$i+3$}&\makebox[1.5em]{$i+4$}&\makebox[1.5em]{$i+5$}\\\hline
$0$ &\color{red}{$s_{i-2}[0]$} &\color{magenta}{$s_{i-1}[0]$} & \color{blue}{$s_i[0]$} &$s_{i+1}[0]$&$s_{i+2}[0]$&$s_{i+3}[0]$ &$s_{i+4}[0]$&$s_{i+5}[0]$\\\hline
$1$ & $s_{i-2}[1]$ &$\color{red}{s_{i-1}[1]}$ & $\color{magenta}{s_i[1]}$ & \color{blue}{$s_{i+1}[1]$}&$s_{i+2}[1]$&$s_{i+3}[1]$ &$s_{i+4}[1]$&$s_{i+5}[1]$\\\hline
$2$ &$s_{i-2}[2]$ &$s_{i-1}[2]$ & $\color{red}{s_i[2]}$ &\color{magenta}{$s_{i+1}[2]$}&\color{blue}{$s_{i+2}[2]$}&$s_{i+3}[2]$ &$s_{i+4}[2]$&$s_{i+5}[2]$\\\hline
$3$ &$\ddots$ & $\ddots$& \color{red}{$s_{i-2}[0]$} &\color{magenta}{$s_{i-1}[0]$} & \color{blue}{$s_i[0]$} &$\ddots$&$\ddots$&$\ddots$\\\hline
$4$ &$\ddots$ &$\ddots$ &$\ddots$  &$\ddots$ &\parbox[c]{0.6 in}{\color{red}{$s_{i-2}[0]\\ +s_{i-1}[1]\\+ s_{i}[2]$}} &  \parbox[c]{0.6 in}{\color{magenta}{$s_{i-1}[0]\\ +s_{i}[1]\\+ s_{i+1}[2]$}}&   \parbox[c]{0.6 in}{\ \color{blue}{$s_i[0] \\+s_{i+1}[1]\\+ s_{i+2}[2]$}} &$\ddots$\\\hline
$5$ &$\ddots$ &$\ddots$ & $\ddots$ &$\ddots$&$\ddots$&\parbox[c]{0.6 in}{\color{red}{$s_{i-1}[1]\\+ 2s_{i}[2]$}}&\parbox[c]{0.6 in}{\color{magenta}{$s_{i}[1]\\+ 2s_{i+1}[2]$}}&\parbox[c]{0.6 in}{\ \color{blue}{$s_{i+1}[1]\\+ 2s_{i+2}[2]$}} \\\hline
\end{tabular}
\centering
\caption{Example of a $(6,3,5)$ convolutional code with $T = 4$ obtained through interleaving a $(6, 3)$ block code.}
\label{fig:exp}
\end{figure}

\section{Construction of the Code}
\label{sec:IV}

In this paper we consider the high rate regime with $R = \frac{k}{n} \geq \frac{1}{2}$. We define $k \triangleq T - N + 1$ and $n \triangleq k + B$ in the same   manner as Fong et al. in \cite{FKLTZA2018}. Note that this implies that $k \ge B$, which will be assumed throughout the paper.
\subsection{Proposed Construction}
\label{sec:hr_overview}
We first introduce two quantities $M$ and $\delta$ as defined below:

\begin{align}
\label{eq:delta}
M = \left\lfloor\frac{B}{N}\right\rfloor \qquad\qquad\qquad\qquad\qquad\qquad \delta = B - N \cdot M, \qquad 0\leq \delta < N.
\end{align}

Note that $\delta$ and $M$ are the remainder and quotient respectively of dividing $B$ by $N$. Our proposed block code which satisfies the set of constraints $(W=T+1, T, B, N)$ is a systematic block code with generator matrix:
\begin{align}
\bG =  \begin{bmatrix} \bI_k & \bP \end{bmatrix}
\label{eq:hr_gen}
\end{align}
where $\bI_k$ is a $k \times k$ identity matrix and $\bP$ is an $k \times B$ parity matrix as defined below:

\begin{align}
\label{matrix:h-rate}
\textbf{P} =
\left[
\renewcommand{\arraystretch}{1.2}
\begin{array}{c}
\begin{array}{c;{2pt/2pt}c}
\mbox{$\textrm{\textbf{P}}_{\textstyle{\delta}}$ \qquad} & \bm{0}^{\textstyle{\delta} \times \scriptstyle{(B-N)}} \qquad\qquad\qquad\qquad\qquad\qquad\qquad\qquad\qquad\qquad
\end{array}
\\ 
\hdashline[2pt/2pt]
\begin{array}{l;{2pt/2pt}c;{2pt/2pt}c}
\bm{0}^{N \times \textstyle{\delta}} & \mbox{$\qquad \textrm{\textbf{P}}_0 $} \qquad \qquad & \bm{0}^{N \times (B-N-\textstyle{\delta}\scriptstyle{)}} \qquad\qquad\qquad\qquad\qquad\qquad\qquad\qquad
\end{array} 
\\
\hdashline[2pt/2pt]
\ddots
\\
\hdashline[2pt/2pt]
\begin{array}{c;{2pt/2pt}c;{2pt/2pt}c}
\qquad \qquad \qquad \qquad \bm{0}^{N \times (\delta + N \cdot (M-2))} & \qquad \mbox{$\textrm{\textbf{P}}_{M-2} $} \qquad \qquad & \bm{0}^{N \times (B - (M-1) \cdot N - \delta)}
\end{array}
\\
\hdashline[2pt/2pt]
\begin{array}{c;{2pt/2pt}c}
\qquad\qquad\qquad\qquad
\qquad\qquad
\bm{0}^{N \times (\textstyle{\delta} \scriptstyle{+ N\cdot (M-1))}} & \qquad \mbox{ $ \textrm{\textbf{P}}_{M-1}$}
\end{array}
\\
\hdashline[2pt/2pt]
\mbox{$\textrm{\textbf{W}}^{(k-B) \times B}$}
\end{array}
\right]
\end{align}
with

\begin{enumerate}
\item \textbf{P}$_{\del} \in \F_{q^m}^{\delta \times N}$ denotes a $\delta \times N$ matrix over the extension field $\F_{q^m}$ that partially overlaps with $\bP_0$.

\item $\bP_0,\ldots, \bP_{M-1} \in \F_q^{N\times N}$  denote $N\times N$ matrices that are mutually non-overlapping and constructed over the base field $\F_q$.


\item \textbf{W}$^{(k-B) \times B}\in \F_{q^m}^{(k-B) \times B}$ denotes a $(k-B) \times B$ matrix over the extension field that overlaps with all $\bP_0,\ldots, \bP_{M-1}$.

\end{enumerate}
Our proposed choice of the matrices $\bP_\delta, \bP_0, \ldots, \bP_{M-1}, \bW$ is as follows. We select
\begin{align}
\bP= \bP_0 = \bP_1 = \cdots = \bP_{M-1} \in \F_q^{N \times N}
\end{align}
such that $\bP$ denotes the parity check part of a $(2N, N)$ MDS block code with a systematic generator matrix: \begin{align}\bG_\mrm{MDS} = \begin{bmatrix} \bI_N & \bP \end{bmatrix}.\end{align}

We further consider a systematic generator matrix associated with a maximum rank distance Gabidulin code $\bG_\mrm{MRD} \in \F_{q^m}^{(k-B+\delta) \times (k+\delta)}$ of the following form:

\begin{align}
\label{eq:Gmrd_hr}
\bG_\mrm{MRD} = \left[ 
\begin{array}{c;{2pt/2pt}c}
\begin{array}{c}
\bI_{k-B+\delta}
\end{array}&
\begin{array}{c}
\bP_A \\\hdashline[2pt/2pt]
\bP_B
\end{array}
\end{array}
\right] \in \F_{q^m}^{(k-B+\delta) \times (k+\delta)},
\end{align}
where $\bI_{k-B+\delta}$ denotes an identity matrix of size $(k-B+\delta)$, $\bP_A$ is a matrix of dimension $\delta  \times B$ and $\bP_B$ is a matrix of dimension $(k-B) \times B$.
By \textit{Lemma 2} such a matrix can be constructed using the Gabidulin construction and transforming the generator matrix into a systematic form. In addition, since we select $m \ge k+\delta$ it also guarantees the code to be MDS.  

In our construction we will select the following:
\begin{align}
&\bW = \bP_B \in \F_{q^m}^{(k-B) \times B},&  \bP_\delta = \begin{bmatrix}\bP_A(1)  & \cdots & \bP_A(N)\end{bmatrix} \in \F_{q^m}^{\delta \times N},
\label{eq:Wselect_HR}
\end{align}
where $\bP_A(1), \cdots, \bP_A(N) $ denote the first $N$ columns of the matrix $\bP_A$.

We present the reader with an example of our construction.

\begin{Example}
Possible \textbf{G} matrix for parameters $(W = 10, T=9, B = 5, N = 3)$

In this example note that $M = \floor{\frac{B}{N}}=1$ and $\delta = B - N \cdot M = 2$. We make the following choices for the constituent matrices:

\begin{align}
\bG_\mrm{MDS} = \left[\begin{array}{c;{2pt/2pt}c}  \bI_3 ~&~  \begin{array}{ccc} \beta_{20} & \beta_{21} & \beta_{22} \\ \beta_{10} &\beta_{11} & \beta_{12}  \\  \beta_{00}&\beta_{01} &\beta_{02} \end{array} \end{array}\right]
\end{align}

\begin{align}
\bG_\mrm{MRD} =  \left[\begin{array}{c;{2pt/2pt}c}  \bI_4 ~&~  
\begin{array}{ccccc} 
\alpha_{30} & \alpha_{31} & \alpha_{32}~ &~ \al_{33} ~&~ \al_{34} \\ 
\alpha_{20} & \alpha_{21} & \alpha_{22}~ &~ \al_{23} ~&~ \al_{24} \\ 
\alpha_{10} & \alpha_{11} & \alpha_{12}~ &~ \al_{13} ~&~ \al_{14} \\ 
\alpha_{00} & \alpha_{01} & \alpha_{02}~ &~ \al_{03} ~&~ \al_{04} \\ 
\end{array} 
\end{array}\right]
\end{align}

Then we  select:
\begin{align}
&\bP_\del = 
\left[
\begin{array}{ccc} 
\alpha_{30} & \alpha_{31} & \alpha_{32} \\
\alpha_{20} & \alpha_{21} & \alpha_{22} \\ 
\end{array} 
\right]
&\bW = \left[
\begin{array}{ccccc}
\alpha_{10} & \alpha_{11} & \alpha_{12}~ &~ \al_{13} ~&~ \al_{14} \\ 
\alpha_{00} & \alpha_{01} & \alpha_{02}~ &~ \al_{03} ~&~ \al_{04} \\
\end{array}\right]
\end{align}
and the resulting generator matrix is given by:
\begin{align}
\label{matrix:example1}
\textbf{G} = 
\left[
\begin{array}{c c c c c c c;{2pt/2pt} c c c c c}
1 & 0 & 0 & 0 & 0 & 0 & 0 & \alpha_{30} & \alpha_{31} & \alpha_{32}& ~0 &~0
\\
0 & 1 & 0 & 0 & 0 & 0 & 0 &\alpha_{20} & \alpha_{21} & \alpha_{22} & ~0 & ~0
\\
0 & 0 & 1 & 0 & 0 & 0 & 0&0 & 0&\beta_{20} &\beta_{21} & \beta_{22}
\\
0 & 0 & 0 & 1 & 0 & 0 & 0&0 & 0&\beta_{10} & \beta_{11} & \beta_{12}
\\
0 & 0 & 0 & 0 & 1 & 0 & 0&0 & 0&\beta_{00} & \beta_{01} & \beta_{02}
\\
0 & 0 & 0 & 0 & 0 & 1 & 0&\alpha_{10} & \alpha_{11} & \alpha_{12}~ &~ \al_{13} ~&~ \al_{14} 
\\
0 & 0 & 0 & 0 & 0 & 0 & 1 &\alpha_{00} & \alpha_{01} & \alpha_{02}~ &~ \al_{03} ~&~ \al_{04} \
\end{array}
\right]
\end{align}
 \end{Example}

\begin{remark}
	Note that in the case when $\delta = 0$, the construction is very similar to the one presented by Martinian and Sundberg in \cite{MartinianSundberg2004} for the case when $N = 1$.	The repetition of the first $B$ symbols in the parity part is replaced by diagonally arranged $N \times N$ matrices consisting of the parity part of an $(2N,N)$ MDS code generator matrix, whereas the MDS code used in \cite{MartinianSundberg2004} is made to be an Gabidulin code over the extension field.

\begin{Example}
Possible \textbf{G} matrix for parameters $(W = 11, T=10, B = 4, N = 2)$.
In this case $M=2$ and $\del=0$. Our proposed construction is as follows:
\end{Example}

\begin{align}
\label{matrix:example2}
\textbf{G} = 
\left[
\begin{array}{c;{2pt/2pt}c}
\begin{array}{c c c c c c c c c}
1 & 0 & 0 & 0 & 0 & 0 & 0 & 0 & 0
\\
0 & 1 & 0 & 0 & 0 & 0 & 0 & 0 & 0
\\
0 & 0 & 1 & 0 & 0 & 0 & 0 & 0 & 0
\\
0 & 0 & 0 & 1 & 0 & 0 & 0 & 0 & 0
\\
0 & 0 & 0 & 0 & 1 & 0 & 0 & 0 & 0
\\
0 & 0 & 0 & 0 & 0 & 1 & 0 & 0 & 0
\\
0 & 0 & 0 & 0 & 0 & 0 & 1 & 0 & 0
\\
0 & 0 & 0 & 0 & 0 & 0 & 0 & 1 & 0
\\
0 & 0 & 0 & 0 & 0 & 0 & 0 & 0 & 1
\end{array}
&
\begin{array}{c}
\begin{array}{c c;{2pt/2pt} c c}
\beta_{10} & \beta_{11} & 0 & 0
\\
\beta_{00} & \beta_{01} & 0 & 0
\\
\hdashline[2pt/2pt]
0 & 0 & \beta_{10}& \beta_{11}
\\
0 & 0 & \beta_{00} & \beta_{01}
\end{array}
\\
\hdashline[2pt/2pt]
\begin{array}{c c c c}
\al_{40} & \al_{41} & \al_{42} & \al_{43}
\\
\al_{30} & \al_{31} & \al_{32} & \al_{33}
\\
\al_{20} & \al_{21} & \al_{22} & \al_{23}
\\
\al_{10} & \al_{11} & \al_{12} & \al_{13}
\\
\al_{00} & \al_{01} & \al_{02} & \al_{03}
\end{array}
\end{array}
\end{array}
\right]
\end{align}

where we use the systematic Gabidulin code:

\begin{align}
\bG_\mrm{MRD} = \left[ \begin{array}{c c} \bI_5 & \begin{array}{c c c c}
\al_{40} & \al_{41} & \al_{42} & \al_{43}
\\
\al_{30} & \al_{31} & \al_{32} & \al_{33}
\\
\al_{20} & \al_{21} & \al_{22} & \al_{23}
\\
\al_{10} & \al_{11} & \al_{12} & \al_{13}
\\
\al_{00} & \al_{01} & \al_{02} & \al_{03}
\end{array}
\end{array} \right]
\end{align}

and MDS code:
\begin{align}
\bG_\mrm{MDS} = \left[ \begin{array}{c c} \bI_2 & \begin{array}{c c}
\beta_{10}& \beta_{11} \\
\beta_{00} & \beta_{01}
\end{array}
\end{array}\right]
\end{align}

as constituent codes. 
\end{remark}

\subsection{Recovery from Arbitrary Erasure Patterns}
\label{subsec:arbitrary}

In order to demonstrate the recovery from arbitrary erasure pattern we will establish the following lemmas.
\begin{Lemma}
\label{lemma:nonurgent_hr}
For any sequence of $N$ erasures in arbitrary positions,  our construction presented in section~\ref{sec:hr_overview} guarantees recovery of the first $\delta$ source symbols $s_0, \ldots, s_{\delta-1}$  and last $k - B$ symbols $s_{B},\ldots, s_{k-1}$ by time $t = T$, i.e., the deadline of the first source symbol $s_0$. 
\end{Lemma}

\begin{Lemma}
\label{lemma:urgent_hr}
For any sequence of $N$ erasures in arbitrary positions, $\forall j \in \{0,1,\dots,M-1\}$, the recovery of the  $s_{\delta + j \cdot N},\ldots, s_{\delta + (j+1) \cdot N - 1}$ is guaranteed by time
$t = T + \delta + j \cdot N$, i.e., the deadline of the first symbol $s_{\delta + j \cdot N}$ in this block of symbols.
\end{Lemma}

\begin{remark}
Lemma~\ref{lemma:nonurgent_hr} and~\ref{lemma:urgent_hr} together establish that every source symbol will be recovered by its respective deadline for any pattern of $N$ erasures in arbitrary positions. In fact the first $\delta$ and last $k-B$ symbols are guaranteed to be recovered by time $t=T$, i.e., the deadline of the first source symbol. In addition the set of the $N$ source symbols associated with the block $\bP_j$ (for $j=0,\ldots, M-1$) in~\eqref{matrix:h-rate} are guaranteed to be recovered by the deadline of the first symbol in this block.
\end{remark}

We will provide a proof of Lemma~\ref{lemma:nonurgent_hr} and~\ref{lemma:urgent_hr} in the remainder of this subsection. 
\subsubsection{Proof of Lemma~\ref{lemma:nonurgent_hr}}

For convenience, we will define the interval of symbols of interest as:
\begin{align}\cI_0 = [0, \delta-1] \cup [B, k] \label{eq:hr_urg_int}\end{align}

We will show that for any pattern of $N$ erasures, the symbols in the interval  $\cI_0$ will be recovered by time $t=T$. As a worst case scenario, we will assume that all erasures are concentrated in the interval $[0,T]$, which is the interval of interest here. This is indeed the most unfavorable scenario given that all the erasures exclusively affect the transmitted symbols that are useful for recovery of the symbols in the interval  $\cI_0$.

Let us denote by $l$ , with $l \in \{0, 1, \dots, N\}$, the number of erasures that affect the remaining middle $N \cdot M$ symbols. By extension $N-l$ the erasures that affect remainder of the interval including source symbols in $t \in \cI_0$. We will consider three distinct cases for $l$ and argue that in each case the statement of Lemma~\ref{lemma:nonurgent_hr} holds.

Before we proceed, note that since $k\triangleq T-N+1$ in our construction the interval $[0,T] = [0,k+N-1]$, i.e., we consider the first $N$  columns following the identity matrix in~\eqref{eq:hr_gen}.

\underline{$l = 0$}:

In this case none of the middle $N \cdot M$ symbols are erased and therefore we can ignore them. To establish the recovery of symbols in $\cI_0$,  in~\eqref{eq:hr_urg_int}, it suffices to consider the following generator matrix associated with~\eqref{eq:hr_gen}.
\begin{align}
\label{matrix:h-rate-1}
\textbf{G}'=
\left[
\renewcommand{\arraystretch}{1.5}
	\begin{array}{c;{2pt/2pt}c}
		\begin{array}{c}
			\begin{array}{c;{2pt/2pt}c}
				\bI_{\delta} & \textbf{0}^{\delta \times (k-B)}
				\\
			\hdashline[2pt/2pt]
				 \textbf{0}^{(k-B) \times \delta} & \bI_{\small{k-B}}
			\end{array}
		\end{array}&
		\begin{array}{c}
			\textrm{\textbf{P}}_{\delta}
			\\ 
			\hdashline[2pt/2pt]
			^{*}\textrm{\textbf{W}}^{(k-B) \times N}
		\end{array}
	\end{array}
\right]
\end{align}
with $^{*}\textrm{\textbf{W}}^{(k-B) \times N}$ a $(k-B) \times N$ matrix consisting of the first $N$ columns of $\textrm{\textbf{W}}^{(k-B) \times B}$. The problem reduces to recovering all source symbols in $\bs_{\cI_0}$ simultaneously from $N$ erasures in arbitrary positions from the received codeword $\bx' = \bs_{\cI_0}  \cdot \bG'$, where $\bs_{\cI_0}$ denotes the source symbols associated with the interval $\cI_0$.  Indeed, such a reduction is immediate since we assume that $l=0$ and thus there are no erasures in the middle $N\cdot M$ symbols. Thus in our construction of $\bP$ in \eqref{matrix:h-rate}, the interference of symbols associated with $\bP_0$  can be immediately cancelled at the decoder, and the effective generator matrix in~\eqref{matrix:h-rate-1} results.

It thus suffices to show that the matrix $\bG'$ in~\eqref{matrix:h-rate-1} corresponds to an MDS code. Note from~\eqref{eq:Gmrd_hr} that the matrix $\bG'$ consists of the first $k-B+\delta+N$ columns of $\bG_\mrm{MRD}$ and hence is an $(k-B+\delta+N, k-B+\delta)$ MRD code over $\F_{q^m}$ by Lemma~\ref{lem:syscol}. Furthermore since we set $m \ge k+\delta$ it follows that $\bG'$ is also a $(k-B+\delta+N, k-B+\delta)$ MDS code. This completes our argument.

\underline{$l \geq N-\delta$}:

In this setting, we have at most $N-l \le N -(N-\delta) = \delta$ erasures affecting the source symbols in the interval $\cI_0$. In this case we claim that all the source symbols of interest in $\cI_0$ will be recovered by time $t = k+\delta-1 < T$. This follows since $\delta < N$ by construction. The effective generator matrix we consider in this case is given by the following:
\begin{align}
\begin{split}
\textbf{G}'=
\left[
\renewcommand{\arraystretch}{1.5}
	\begin{array}{c;{2pt/2pt}c}
			\begin{array}{c;{2pt/2pt}c}
				\bI_{\small{\delta}} &\textbf{0}^{\delta \times (k-B)}
			\\ 
			\hdashline[2pt/2pt]
				 \textbf{0}^{(k-B) \times \delta} & \bI_{\small{k-B}}
			\end{array}
		&
		\begin{array}{c}
			\textrm{\textbf{P}}_{\delta -}
			\\ 
			\hdashline[2pt/2pt]
			\textrm{\textbf{W}}_1^{(k-B) \times \delta}
		\end{array}
	\end{array}
\right]
\end{split}
\label{eq:gen_eff_2}
\end{align}
with
\begin{enumerate}
\item \textbf{P}$_{\delta - }$ a $\delta \times \delta$ matrix corresponding to the first $\delta$ columns of \textbf{P}$_{\delta}$.
\item \textbf{W}$_1^{(k-B) \times \delta}$ a $(k-B) \times \delta$ matrix corresponding to the first $\delta$ columns of \textbf{W}$^{(k-B) \times B}$.
\end{enumerate}

Note that the matrix in~\eqref{eq:gen_eff_2} is obtained from the first $\delta$ columns following the systematic part in the generator matrix $\bG$ in~\eqref{eq:hr_gen}. In these columns, there is no interference from the middle $N\cdot M$ symbols associated with $\bP_0,\ldots, \bP_{M-1}$ (c.f.~\eqref{matrix:h-rate}). Thus in the recovery of symbols in $\cI_0$, it suffices to show that in the codeword $\bx' = \bs_{\cI_0} \cdot \bG'$, we can recover all the source symbols $\bs_{\cI_0}$ from an arbitrary pattern of $\delta$ erasures.  This claim however follows by noting that the matrix $\bG'$ in~\eqref{eq:gen_eff_2} consists of the first $k-B+2\delta$ columns of $\bG_\mrm{MRD}$ 
defined in~\eqref{eq:Gmrd_hr}. Thus  $\bG'$ is a  $(k-B+2\del, k-B+\del)$ MRD code over $\F_{q^m}$ by Lemma~\ref{lem:syscol}. Furthermore since $m \ge k+\del > k-B+2\del$ it follows that $\bG'$ is also a $(k-B+2\del, k-B+\del)$ MDS code. This completes the argument for the case when $l \ge N-\delta$

\underline{$1 \leq l < N - \delta$}:


This case if the most challenging case as we have to consider the overlap between $\bP_0$ and $\bP_\delta$ in~\eqref{matrix:h-rate}. We  consider the  submatrix $\bG'$ of $\bG$ in~\eqref{eq:hr_gen}, consisting of its first $k + N$ columns in the interval $[0,T]$.  
\begin{align}
\label{matrix:G'1}
\bG' =
\left[
\renewcommand{\arraystretch}{1.5}
\begin{array}{c;{2pt/2pt}c}
	\begin{array}{c;{2pt/2pt}c;{2pt/2pt}c}
		\scriptscriptstyle \bI_{\delta} & \scriptstyle \bm{0}^{\delta \times N} &  \scriptstyle \bm{0}^{\delta \times (k-B)}
		\\ 
		\hdashline[2pt/2pt]
		\scriptstyle \bm{0}^{N \times \delta} & \scriptscriptstyle \bI_{N} &  \scriptstyle \bm{0}^{N \times (k-B)}
		\\ 
		\hdashline[2pt/2pt]
		 \scriptstyle \bm{0}^{(k-B) \times (B-N)} &  \scriptstyle \bm{0}^{(k-B) \times N} & \scriptscriptstyle \bI_{k-B} 
	\end{array}
&
	\begin{array}{c;{2pt/2pt}c}
		\textrm{\small\textbf{P}}_{\delta \scriptscriptstyle{-}}
		&
		\textrm{\small\textbf{P}}_{\delta \scriptscriptstyle {+}}
	\\
	\hdashline[2pt/2pt]
	\scriptstyle \bm{0}^{\scriptscriptstyle N \times \delta}
	&
	\textrm{\small*\textbf{P}}_0^{\scriptscriptstyle N \times (N-\delta)}
	\\ 
	\hdashline[2pt/2pt]
	\textrm{\small\textbf{W}}_1^{\scriptscriptstyle (k-B) \times \delta}
	&
	\textrm{\small\textbf{W}}_2^{\scriptscriptstyle (k-B) \times (N-\delta)}
	\end{array}
\end{array}
\right]
\end{align}
with
\begin{enumerate}
\item \textbf{P}$_{\delta \scriptscriptstyle{-}}$ a $\delta \times \delta$ matrix corresponding to the first $\delta$ columns of \textbf{P}$_{\delta}$.
\item \textbf{P}$_{\delta \scriptscriptstyle{+}}$ is a $\del \times (N-\del)$ matrix corresponding to the last $N - \delta$ columns of \textbf{P}$_{\delta}$.
\item \textrm{\small*\textbf{P}}$_0^{\scriptscriptstyle N \times (N-\delta)}$ a $ N \times (N-\delta)$ matrix corresponding to the first $N - \delta$ columns of \textbf{P}$_0$.
\item \textbf{W}$_1^{(k-B) \times \delta}$ a $(k-B) \times \delta$ matrix corresponding to the first $\delta$ columns of \textbf{W}$^{(k-B) \times B}$.
\item \textbf{W}$_2^{(k-B) \times (N-\delta)}$ a $(k-B) \times (N-\delta$) matrix corresponding to the last $N-\del$ columns of \textbf{W}$^{(k-B) \times B}$.
\end{enumerate}

In our expression for $\bG'$ in~\eqref{matrix:G'1},   the all-zero rows as well as the identity matrices corresponding to $\bP_1, \ldots \bP_M$ are removed as they do not interfere with the parity symbols in
the interval $[0,\delta+N-1] \cup [B, T]$. Note that the code associated with the generator matrix $\bG'$  is equivalent to truncating the codewords associated with $\bG$ to the interval $[0,T]$. Thus the  codeword in the interval $[0,T]$ can be expressed as 
\begin{align}
\bx' = \begin{bmatrix}\bs_\delta & \bs_0 & \bs_L \end{bmatrix} \cdot \bG' = \begin{bmatrix} \bs_\delta & \bs_0 & \bs_L & \bc \end{bmatrix}, \label{eq:xp_eqn}
\end{align}
where $\bs_\del = [s_0, \ldots s_{\del-1}]$ and $\bs_L = [s_{k-B+1}, \ldots, s_{k-1}]$ denote the first $\del$ and the last $k-B$ source symbols respectively, 
while $\bs_0 = [s_{\del}, \ldots, s_{\del+N-1}]$ represents the interfering source symbols that are associated with $\bP_0$.  The second equality follows since $\bG'$ is systematic and by selecting:
\begin{align}
\bc = \begin{bmatrix}\bs_\delta & \bs_0 & \bs_L \end{bmatrix}
\cdot 
\left[
\renewcommand{\arraystretch}{1.5}
\begin{array}{c}
	\begin{array}{c;{2pt/2pt}c}
		\textrm{\small\textbf{P}}_{\delta \scriptscriptstyle{-}}
		&
		\textrm{\small\textbf{P}}_{\delta \scriptscriptstyle {+}}
	\\
	\hdashline[2pt/2pt]
	\scriptstyle \bm{0}^{\scriptscriptstyle N \times \delta}
	&
	\textrm{\small*\textbf{P}}_0^{\scriptscriptstyle N \times (N-\delta)}
	\\ 
	\hdashline[2pt/2pt]
	\textrm{\small\textbf{W}}_1^{\scriptscriptstyle (k-B) \times \delta}
	&
	\textrm{\small\textbf{W}}_2^{\scriptscriptstyle (k-B) \times (N-\delta)}
	\end{array}
\end{array}
\right] \in \F_{q^m}^{N}
\end{align}
as the first $N$ parity symbols of the codeword. Note that unlike the previous cases one has to explicitly account for the interference
from the symbols in $\bs_0$. We let 
\begin{align}
\bx_1' = \begin{bmatrix} \bs_\del & \bs_L & \bc\end{bmatrix} \label{eq:aux_x_def}
\end{align}
denote the portion of the codeword $\bx'$ not involving $\bs_0$ and observe from~\eqref{matrix:G'1} and~\eqref{eq:xp_eqn} that we can express:
\begin{align}
\bx_1' = \begin{bmatrix}\bs_\del & \bs_L \end{bmatrix} \bG'_\mrm{MRD} + \bs_0  \cdot\bT, \label{eq:bx1_hr_eqn}
\end{align}
where the matrices $\bG'_\mrm{MRD}$ and $\bT$ can be expressed as:
\begin{align}
\bG'_\mrm{MRD} =
\left[
\renewcommand{\arraystretch}{1.5}
	\begin{array}{c;{2pt/2pt}c}
		\begin{array}{c}
			\begin{array}{c;{2pt/2pt}c}
				\bI_{\delta} & \textbf{0}^{\delta \times (k-B)}
				\\
			\hdashline[2pt/2pt]
				 \textbf{0}^{(k-B) \times \delta} & \bI_{\small{k-B}}
			\end{array}
		\end{array}&
		\begin{array}{c}
			\textrm{\textbf{P}}_{\delta}
			\\ 
			\hdashline[2pt/2pt]
			^{*}\textrm{\textbf{W}}^{(k-B) \times N}
		\end{array}
	\end{array}
\right] \in \F_{q^m}^{(\del+k-B)\times (\del+k-B+N)}
\end{align}

and
\begin{align}
\bT = \begin{bmatrix} {\bf 0}^{N \times (2\del + k-B+N)}  & 	\textrm{\small*\textbf{P}}_0^{\scriptscriptstyle N \times (N-\delta)}
 \end{bmatrix}
 \in \F_{q}^{N \times (k-B+N+\del)}
\end{align}

Note that by construction $\bG'_\mrm{MRD}$ is a generator matrix of a $(k-B+\del+N, k-B+\del)$ Gabidulin code (in systematic form) over $\F_{q^m}$
while $\bT$ is a matrix over the base field $\F_q$. Further if we let $\{\bT(0),\ldots, \bT(N-1)\}$ denote the rows of the matrix $\bT$, note that:
\begin{align}
\bs_0 \cdot \bT  = \sum_{i=0}^{N-1} s_{\del+i} \cdot \bT(i) \label{eq:sum_hr_interf}
\end{align}

Recall that there are at most $l$ erasures among the $N$ symbols in $\bs_0$. We will  subtract the contribution of the remaining $N-l$ symbols in the right hand side of~\eqref{eq:sum_hr_interf} (and equivalently~\eqref{eq:bx1_hr_eqn}). Let $\tilde{\bs}_0$ denote the erased symbols in $\bs_0$ and $\tilde{\bT} \in \F_q^{l \times (k-B+N+\del)}$ a matrix whose rows corresponds to the erased symbols in $\bs_0$. With the cancellation of the non-erased symbols in $\bs_0$, we can rewrite the last term in~\eqref{eq:bx1_hr_eqn} as the following:
\begin{align}
\bx_2' = \begin{bmatrix}\bs_\del & \bs_L \end{bmatrix} \cdot  \bG'_\mrm{MRD} + \tilde{\bs}_0  \cdot \tilde{\bT}. \label{eq:bx2_hr_eqn}
\end{align}

Recall that there can be at most $N-l$ erasures in  $\bx_2'$. If we let $\hat{\bx}_2'$ to denote the non-erased locations in $\bx_2'$ we note that $\hat{\bx}_2'$ is of size at-least 
$$(k-B+\del+N) - (N-l) = k-B+\del+l.$$
We assume, as a worst case, that the size of $\hat{\bx}_2'$ is equal to $k-B+\del+l$.
We let $\hat{\bG}'_\mrm{MRD}\in \F_{q^m}^{(k-B+\del)\times (k-B+\del+l)}$ to denote a submatrix of ${\bG}'_\mrm{MRD} $ whose columns correspond to the non-erased position in $\hat{\bx}_2'$
and likewise let $\hat{\bT}\in \F_{q}^{l \times (k-B+\del+l)}$ denote a submatrix of $\tilde{\bT}$ whose columns correspond to the non-erased symbols in $\hat{\bx}_2'$.
Thus~\eqref{eq:bx2_hr_eqn} can be reduced to:
\begin{align}
\hat{\bx}_2' = \begin{bmatrix}\bs_\del & \bs_L \end{bmatrix} 
\cdot \hat{\bG}'_\mrm{MRD} + \tilde{\bs}_0  \cdot \hat{\bT}. \label{eq:hx2_hr_eqn}
\end{align}
Note that al the symbols in $\hat{\bx}_2' $ are received at the decoder by time $t=T$. Thus it remains to argue that the receiver can uniquely recover $\{\bs_\del,~\bs_L\}$
from~\eqref{eq:hx2_hr_eqn}. Towards this end note that $\hat{\bT} \in \F_{q}^{l \times (k-B+\del+l)}$ is a matrix in $\F_q$ of rank no more than $l$. Thus the null-space of $\hat{\bT}$
has a rank of at least $k-B+\del$, i.e., there exists a matrix $\bM \in \F_q^{(k-B+\del+l)\times (k-B+\del)}$ with full rank such that
\begin{align}
\hat{\bT} \cdot \bM = {\bf 0}^{l\times (k-B+\del)}. \label{eq:cancel_hr}
\end{align}
Multiplying both sides of~\eqref{eq:hx2_hr_eqn} by $\bM$ and using~\eqref{eq:cancel_hr} we have that:
\begin{align}
\hat{\bx}_2' \cdot \bM = \begin{bmatrix}\bs_\del & \bs_L \end{bmatrix} 
\cdot \hat{\bG}'_\mrm{MRD} \cdot\bM.
\end{align}

Thus it only remains to show that the matrix $\hat{\bG}'_\mrm{MRD} \cdot\bM \in \F_{q^m}^{(k-B+\del) \times (k-B+\del)}$ is a full-rank matrix. However since $\bM \in  \F_q^{(k-B+\del+l)\times (k-B+\del)}$ is a full-rank matrix in the base field and $\hat{\bG}'_\mrm{MRD} $ corresponds to the generator matrix of a $(k-B+\del+l, k-B+\del)$ MRD code by Lemma~\ref{lem:syscol}, the product is guaranteed to be a full-rank square matrix by Theorem 1. Thus one can uniquely recover $\{\bs_\del,~\bs_L\}$ as required.

This completes the proof of recovery of the first $\del$ and the last $k-B$ symbols by time $t=T$ as required.

$\hfill\blacksquare$

\begin{remark}
The system of equations in~\eqref{eq:hx2_hr_eqn} is closely related to the error recovery problem studied in~\cite{Sil08}. Indeed one could view $\tilde{\bs}_0 \in \F_{q^m}^l$ 
as $l$ errors introduced in the transmission of a codeword from a $(k-B+\del+l, k-B+\del)$ MRD code. The result in~\cite{Sil08} guarantees that only half of the rank-distance, i.e., $l/2$ errors can be corrected. In contrast we are able to correct up to $l$ errors. The main difference with~\cite{Sil08} is that we assume that the matrix $\hat{\bT}$  is known to the receiver while the result in~\cite{Sil08} assumes that such a matrix is not available at the receiver. 
\end{remark}

\begin{remark}
We do not guarantee the recovery of the symbols in $\bs_0$ by time $t=T$. Indeed depending upon the nature of erasure patterns the matrix $\tilde{\bT}$ in~\eqref{eq:bx2_hr_eqn} may not be full rank and the $l$ erased symbols in $\bs_0$ may  not be recovered by time $t=T$.  However the first $\del$ and the last $k-B$ symbols are guaranteed to be recovered for any pattern of $N$ erasures by time $t=T$.
\end{remark}

\subsubsection{Proof of Lemma~\ref{lemma:urgent_hr}}
From Lemma~\ref{lemma:nonurgent_hr} the symbols in the interval $\cI_0 = [0,\del-1] \cup [B, k-1] $ are guaranteed to be recovered for any pattern of $N$ erasures by time $t=T$.  Upon canceling their contribution in the received codeword  it suffices to consider the following submatrix $\bG'$ of \eqref{matrix:h-rate-1}.
\begin{align}
\label{matrix:G'-2}
\bG'=
\left[
\renewcommand{\arraystretch}{1.2}
\begin{array}{c;{2pt/2pt}c}
\bI_{N\cdot M}
&
\begin{array}{c}
\begin{array}{c;{2pt/2pt}c}
\mbox{$\textrm{\textbf{P}}_{0}$} ~~~~ & ~~~~~~~~~~~~ \bm{0}^{N \times (N \cdot (M-1))}
\end{array}
\\ 
\hdashline[2pt/2pt]
\begin{array}{c;{2pt/2pt}c;{2pt/2pt}c}
\bm{0}^{N \times N} & \mbox{~~~~$\textrm{\textbf{P}}_{1}$~~~} & \bm{0}^{N \times (N \cdot (M-2))}
\end{array}  
\\
\hdashline[2pt/2pt]
\mbox{\dots}
\\
\hdashline[2pt/2pt]
\begin{array}{c;{2pt/2pt}c}
\bm{0}^{N \times (N \cdot (M-1))} ~~~~~~~~~~~~~& \mbox{$~~~ \textrm{\textbf{P}}_{M-1}$}
\end{array}
\end{array}
\end{array}
\right]
\end{align}

The unnecessary all \textbf{0} matrices where removed w.l.o.g. and for readability concerns. 


As all the blocks $\bP_j$ are mutually non-overlapping the $N$ symbols $\bs_j =\{s_{\del + j\cdot N},\ldots, s_{\del+(j+1)\cdot N -1}\}$  associated with $\bP_j$ are not interfered by any other symbols. The recovery of the symbols $\bs_j$ can therefore be considered with the help of the following submatrix of \eqref{matrix:G'-2}.
\begin{align}
\label{matrix:G'-j}
\textbf{G}'_j = \begin{bmatrix} \bI_N & \bP_j\end{bmatrix} \in \F_q^{N\times N}
\end{align}
Since by construction $\bG'$ is a generator matrix of a $(2N, N)$ MDS code, it is able to recover from any $N$ erasures in arbitrary locations at the end of the block. 
Thus it only suffices to show that the delay constraints are satisfied. The oldest symbol in $\bs_j$, i.e., $s_{\del+j \cdot N}$ must be decoded by time:
\begin{align*} 
t &= \del+j \cdot N + T \\
&=  \del +j \cdot N + k+N-1 \\
&=  \del + (j+1)\cdot N + k -1
\end{align*}

Indeed the  structure of the generator matrix defined in \eqref{matrix:h-rate}, guarantees that the index of the last column of \textbf{P}$_{j}$ corresponds to 
$\del + (j+1) \cdot N + k -1$ as required.

$\hfill\blacksquare$
\subsection{Recovery from Burst Erasure Pattern}

In this section we will show that for any burst erasure of length up to $B$ each source symbol can be recovered by its respective deadline. We will first prove the following lemma for the special case of first $\del$ source symbols.
\begin{Lemma}
For any burst erasure of maximum length $B$ any source symbol in the interval $[0,\del-1]$ is recovered by time $t = k+\del \le T$, i.e., by the deadline of the first symbol $s_0$.
\label{lemma:burst_1}
\end{Lemma}

\begin{IEEEproof}
Let us define $i$ as the index of the beginning of the burst of size $B$.  
Hence the burst spans over the interval $[i, i+B-1]$.
Clearly we must assume that $i \in [0,\del-1]$, otherwise none of the first $\del$ symbols are erased.  Also note that since $B= \del + N\cdot M$, any burst erasure
of length $B$ that starts at position $i$ will erase all the $N\cdot M$ middle symbols and an additional $i$ symbols in the interval $[B, k]$. The total number of erased symbols in the interval $\cI_0 = [0,\del-1] \cup [B,k-1]$ is hence given by $\del-i + i = \del$.  
For the recovery of the symbols in $\cI_0$ we will only consider the first $\del$ parity columns in the generator matrix $\bG$ defined in~\eqref{eq:hr_gen} and \eqref{matrix:h-rate-1}. 
Since there is no interference from the middle $N\cdot M$ symbols in this period our effective generator matrix is given by (c.f.~\eqref{eq:gen_eff_2}):
\begin{align}
\begin{split}
\bG' =
\left[
\renewcommand{\arraystretch}{1.5}
	\begin{array}{c;{2pt/2pt}c}
			\begin{array}{c;{2pt/2pt}c}
				I_{\small{\delta}} &\textbf{0}^{\delta \times (k-B)}
			\\ 
			\hdashline[2pt/2pt]
				 \textbf{0}^{(k-B) \times \delta} & I_{\small{k-B}}
			\end{array}
		&
		\begin{array}{c}
			\textrm{\textbf{P}}_{\delta -}
			\\ 
			\hdashline[2pt/2pt]
			\textrm{\textbf{W}}_1^{(k-B) \times \delta}
		\end{array}
	\end{array}
\right]
\end{split}
\label{eq:gen_eff_2x}
\end{align}
Since $\bG'$ in~\eqref{eq:gen_eff_2x} constitutes a $(k-B+2\del, k-B+\del)$ MDS code, as discussed before, any sequence of $\del$ erasures are guaranteed to be recovered. In particular the erased symbols in the interval $[0,\del-1]$ are all recovered by time $t = k+ \del \le T$ as required.
\end{IEEEproof}

\begin{Lemma}
For any burst erasure of maximum length $B$ and each $j \in \{0,\ldots, M-1\}$, all the $N$ source symbols associated with block $\bP_j$, i.e., $\bs_j = \{s_{\del + j\cdot N},\ldots, s_{\del+(j+1)\cdot N-1}\}$
are recovered by time $t = \del + j\cdot N + T$, i.e., by the deadline of the first symbol $s_{\del + j \cdot N}$. 
\label{lemma:burst_2}
\end{Lemma}
\begin{IEEEproof}
Since the symbols $s_0,\ldots,s_{\del-1}$ are guaranteed to be recovered before time $t=T$ as per Lemma~\ref{lemma:burst_1}, one can cancel their contributions from the received codeword and consider the following generator matrix associated with~\eqref{eq:hr_gen}:
\begin{align}
\label{matrix:G'-burst}
\textbf{G}' =
\left[
\renewcommand{\arraystretch}{1.2}
\begin{array}{c;{2pt/2pt}c}
\bI_{k-\delta}
&
\begin{array}{c}
\begin{array}{c;{2pt/2pt}c}
\mbox{$\textrm{\textbf{P}}_{0}$} ~~~~ & ~~~~~~~~~ \bm{0}^{N \times ((M-1) \cdot N)}
\end{array}
\\ 
\hdashline[2pt/2pt]
\begin{array}{c;{2pt/2pt}c;{2pt/2pt}c}
\bm{0}^{N \times N} & \mbox{~~$\textrm{\textbf{P}}_{1}$~~} & \bm{0}^{N \times ((M-2) \cdot N)}
\end{array}  
\\
\hdashline[2pt/2pt]
\mbox{\dots}
\\
\hdashline[2pt/2pt]
\begin{array}{c;{2pt/2pt}c}
\bm{0}^{N \times ((M-1) \cdot N)} ~~~~~~~~~~~~~& \mbox{$~~~ \textrm{\textbf{P}}_{M-1}$}
\end{array}
\\
\hdashline[2pt/2pt]
\begin{array}{c}
^*\textrm{\textbf{W}}^{(k-B)\times (B-\delta)}
\end{array}
\end{array}
\end{array}
\right]
\end{align}
with $^*$\textbf{W}$^{(k-B) \times (B - \delta)}$ the $(k-B) \times (B - \delta)$ submatrix of \textbf{W}$^{(k-B) \times B}$ consisting of its last $B-\delta$ columns.

We further assume that the burst starts in the block $\cI_j = [\del + j\cdot N, \del + (j+1)\cdot N-1]$ and let $l\in \{0,\ldots, N-1\}$ denote the 
relative position in $\cI_j$ where the burst begins.  It suffices to show that the erased symbols $s_{l + \del + j\cdot N}, \ldots, s_{\del + (j+1) \cdot N-1}$ are recovered
at the end of the last column in block $\bP_j$, which as seen before corresponds to the time $t = \del+(j+1)\cdot N-1$, as required. When treating subsequent blocks, i.e., $j+1, \ldots, M-1$,
the same argument can be repeated (along with $l=0$), as the length of the erasure burst to be considered will only be reduced. 

Since the burst of length $B$ begins at time $t = j\cdot N+\del + l$, it erases the last $N-l$ symbols in the block $\bs_j$ and terminates at time 
$$t_1 = j\cdot N+\del +l + B-1.$$
We will  recover the erased symbols in $\bs_j$ using the last column in $\bP_j$, i.e., by time 
$$t_2 = k+\del + (j+1)\cdot N-1 = T + j\cdot N + \del,$$  where we use the fact that $k \triangleq T+N-1$.
Since all the erasures are consecutive, the number of non-erased symbols following the burst is in the interval of length  $[t_1+1, t_2]$, which is of length
\begin{align}
t_2 - t_1 = k-B+N-l.
\label{eq:interval_win}
\end{align}
We will show that in addition to the $N-l$ erased symbols in  $\bs_j$, the last $k-B$ source symbols $s_{B},\ldots, s_{k-1}$ associated with $\bW^{(k-B)\times B}$ will also be 
simultaneously recovered at the end of block $\bP_j$.
In order to do so we will consider two matrices representing the two possible cases that can arise given the start index of the burst $l$.

\underline{$l \geq k - B \rightarrow N \geq k - B + N - l$ }: 

In this case, the last $k-B+N-l$ columns are entirely contained within the $N$ columns of $\bP_j$.
Upon removing the contribution of the un-erased symbols before the burst, it suffices to consider the following submatrix \textbf{G}$_j$ of \eqref{matrix:G'-burst}. 

\begin{align}
\label{matrix:G'-j1}
\textbf{G}'_j =
\left[
\renewcommand{\arraystretch}{1.5}
	\begin{array}{c;{2pt/2pt}c}
			\begin{array}{c;{2pt/2pt}c}
				\bI_{N-l} & \textbf{0}^{(N-l) \times (k-B)}
			\\ 
			\hdashline[2pt/2pt]
				 \textbf{0}^{(k-B) \times (N-l)} & \bI_{\small{k-B}}
		\end{array}&
		\begin{array}{c}
			^{*}\textrm{\textbf{P}}_{j}^{(N-l) \times (k-B+N-l)}
			\\ 
			\hdashline[2pt/2pt]
			^{*}\textrm{\textbf{W}}^{(k-B) \times (k-B+N-l)}
		\end{array}
	\end{array}
\right]
\end{align}
with
\begin{enumerate}
\item $^{*}$\textbf{P}$_{j}^{(N-l) \times (k-B+N-l)}$ the $(N-l) \times (k-B+N-l)$ submatrix of \textbf{P}$_{j}$ consisting of its last $N-l$ rows and last $k-B+N-l$ columns.
\item $^{*}$\textbf{W}$^{(k-B) \times (k-B+N-l)}$ the $(k-B) \times (k-B+N-l)$ submatrix of \textbf{W}$^{(k-B) \times B}$ from \eqref{matrix:h-rate} consisting of its $[\delta + (j+1) \cdot N - (k - B + N - l), \delta + (j+1) \cdot N -1]$ columns.
\end{enumerate}

The $\bI_{N-l}$ matrix corresponds to the last $N-l$ symbols of the subblock $j$ which are surely erased by the burst. We can also assume (as a worst case assumption) that all symbols of $\bI_{k-B}$ are erased. However the symbols
corresponding to the parity columns in~\eqref{matrix:G'-j1} are not erased as they occur after the burst. 

Let
$\tilde{\bs_j}$ denote the last $N-l$ erased symbols in $\bs_j = \{ s_{\del + j\cdot N}, \ldots, s_{\del + (j+1)\cdot N-1}\}$ and $\bs_L = \{s_B, \ldots, s_{k-1}\}$ denote the last $k-B$  symbols. After canceling the effect of all
symbols before $\tilde{\bs}_j$ the receiver can compute
\begin{align}
\tilde{\bx} = \tilde{\bs}_j  \cdot ^{*}\textrm{\textbf{P}}_{j}^{(N-l) \times (k-B+N-l)} + \bs_L \cdot ^{*}\textrm{\textbf{W}}^{(k-B) \times (k-B+N-l)}
\label{eq:bxt_eqn_HR}
\end{align}
Our decoder will proceed as follows:
\begin{itemize}
\item (Step 1): Null out the matrix $^{*}\textrm{\textbf{P}}_{j}^{(N-l) \times (k-B+N-l)}$, which is over the base field,  and recover $\bs_L$.
\item (Step 2): Cancel the term involving $\bs_L$ in~\eqref{eq:bxt_eqn_HR} and recover  $\tilde{\bs}_j$.
\end{itemize}
Recall that the matrix $\bW^{(k-B) \times N}$ is part of a Gabidulin code (in systematic form) with generator matrix $\bG_\mrm{MRD}$ (c.f.~\eqref{eq:Gmrd_hr},~\eqref{eq:Wselect_HR}) and hence we can express
\begin{align}
\label{matrix:G_MRD-Gab}
\bG_\mrm{MRD} = 
\left[ 
\begin{array}{c;{2pt/2pt}c}
\begin{array}{c}
\bI_{k-B+\delta}
\end{array}&
\begin{array}{c}
\bP_A^{\del \times B} \\\hdashline[2pt/2pt]
\bW^{(k-B) \times B}\end{array}
\end{array}
\right] =
\left[ 
\begin{array}{c;{2pt/2pt}c}
\begin{array}{c}
\bI_{k-B+\delta}
\end{array}&
\begin{array}{c;{2pt/2pt} c;{2pt/2pt} c}
\bP_A^- & ^{*}\bP_A^{\del \times (k-B+N-l)} & \bP_A^+ \\\hdashline[2pt/2pt]
\bW^- & ^{*}\textrm{\textbf{W}}^{(k-B) \times (k-B+N-l)}& \bW^+
\end{array}
\end{array}
\right]
\end{align}
Here $ ^{*}\bP_A^{\del \times (k-B+N-l)} $ is a submatrix of $\bP_A$ whose columns are aligned with  $ ^{*}\textrm{\textbf{W}}^{(k-B) \times (k-B+N-l)}$.
It thus follows that
\begin{align}
\bG_\mrm{MRD}'=\left[ 
\begin{array}{c;{2pt/2pt}c}
\begin{array}{c}
	\bI_{\del} 
	\\
	\hdashline[2pt/2pt]
	{\bf 0}^{(k-B) \times \del}
\end{array}
&
\begin{array}{c}
^{*}\bP_A^{\del \times (k-B+N-l)}
\\
\hdashline[2pt/2pt]
^{*}\textrm{\textbf{W}}^{(k-B) \times (k-B+N-l)}
\end{array}
\end{array}
\right] 
\in \F_{q^m}^{(k-B+\del) \times (k-B+ N-l +\del)}
\end{align}

is the generator matrix of a $(k-B+N-l + \del, k-B+\del)$ MRD code by Lemma~\ref{lem:syscol}.
Since the symbols $\bs_\del$ have already been recovered, the receiver can compute (via~\eqref{eq:bxt_eqn_HR}):
\begin{align}
\hat{\bx} &= \begin{bmatrix} \bs_\del & \bs_\del ^{*} \cdot \bP_A^{\del \times (k-B+N-l)} + \tilde{\bx}\end{bmatrix} \\
&=  \tilde{\bs}_j \cdot ^{*}\hat{\textbf{P}}_{j}^{(N-l) \times (k-B+N-l+\del)} + \begin{bmatrix} \bs_\del & \bs_L \end{bmatrix} \cdot \bG'_\mrm{MRD},
\label{eq:x_expanded_hr1}
\end{align}
where we select
\begin{align}
^{*}\hat{\textbf{P}}_{j}^{(N-l) \times (k-B+N-l+\del)} = \begin{bmatrix} {\bf 0}^{N-l \times \del} &  ^{*}\textrm{\textbf{P}}_{j}^{(N-l) \times (k-B+N-l)}\end{bmatrix}.
\end{align}

Since the rank of $^{*}\hat{\textbf{P}}_{j}^{(N-l) \times (k-B+N-l+\del)} $ is no more than $N-l$ we are guaranteed that there exists a full rank matrix $\bM \in \F_q^{(k-B+N-l+\del) \times (k-B+\del)}$
such that
\begin{align}
^{*}\hat{\textbf{P}}_{j}^{(N-l) \times (k-B+N-l+\del)}\cdot \bM^{(k-B+N-l+\del) \times (k-B+\del)} = {\bf 0}
\label{eq:cancel_M_hr2}
\end{align}

Multiplying both sides of~\eqref{eq:x_expanded_hr1} by the matrix $ \bM^{(k-B+N-l+\del) \times (k-B+\del)}$ from the right and using~\eqref{eq:cancel_M_hr2} we obtain:
\begin{align}
\hat{\bx} \cdot \bM = \begin{bmatrix} \bs_\del & \bs_L \end{bmatrix}
\cdot \bG'_\mrm{MRD} \cdot \bM.
\end{align}

Furthermore since $\bG'_\mrm{MRD}$ is shown to be the generator matrix of a $(k-B+N-l + \del, k-B+\del)$ MRD code and $\bM^{(k-B+N-l+\del) \times (k-B+\del)}$ is a full rank matrix over the base field $\F_{q}$, the product
$\bG'_\mrm{MRD} \cdot \bM$ is a square invertible matrix over $\F_{q^m}$ by Theorem 1. Thus one can uniquely recover $[\bs_\del,~\bs_L]$ and in particular the vector $\bs_L$ as required.

Upon the successful recovery of $\bs_L$, the receiver can cancel its contribution for the right hand side in~\eqref{eq:bxt_eqn_HR}.
In turn we must reconstruct the vector $\tilde{\bs}_j$ from the reduced vector
$$\tilde{\bs}_j \cdot ^{*}\textrm{\textbf{P}}_{j}^{(N-l) \times (k-B+N-l)}$$

However since $\bP_j$ is a Cauchy matrix in $\bG_\mrm{MDS} = \begin{bmatrix}\bI_N & \bP_j \end{bmatrix}$ and  $^{*}\textrm{\textbf{P}}_{j}^{(N-l) \times (k-B+N-l)}$ is a submatrix of $\bP_j$,
every square submatrix of $^{*}\textrm{\textbf{P}}_{j}$ is non-singular given that any submatrix of a Cauchy matrix is invertible. Thus the recovery of $\tilde{\bs}_j$ follows.

\underline{$l < k - B \rightarrow N < k - B + N - l$ }:

In this case, the length-$B$ burst does not erase parity symbols corresponding to any column of \textbf{P}$_j$.

In addition, we need to consider separately two additional sub-cases for this setting. The first concerns a burst erasure starting in the interval $[\delta,\delta+N-1]$ corresponding to the source symbols of the first subblock \textbf{P}$_0$. The second concerns a burst erasure which starts in the interval $[\delta+N,B-1]$ corresponding to the source symbols of the remaining subblocks $\bP_1, \bP_2, \dots, \bP_{M-1}$.

We first present the reader with the more general case 
when the burst starts in the inverval $[\delta+N,B-1]$, the setting can be interpreted with the help of the following submatrix of \eqref{matrix:G'-burst}, $\bG'_j$ for $j \in \{1,2,\dots,M-1\}$ .

\begin{align}
\label{matrix:G'j-B2}
\textbf{G}'_j =
\left[
\renewcommand{\arraystretch}{1.5}
	\begin{array}{c;{2pt/2pt}c}
			\begin{array}{c;{2pt/2pt}c}
				\bI_{N-l} & \textbf{0}^{(N-l) \times (k-B)}
			\\ 
			\hdashline[2pt/2pt]
				 \textbf{0}^{(k-B) \times (N-l)} & \bI_{\small{k-B}}		\end{array}&
		\begin{array}{c;{2pt/2pt}c}
			\bm{0}^{(N-l)\times (k-B-l)} 
			&
			^{*}\textrm{\textbf{P}}_{j}^{(N-l) \times N}
			\\ 
			\hdashline[2pt/2pt]		
			\textrm{\textbf{W}}_1^{(k-B) \times (k-B-l)}	
			&
			\textrm{\textbf{W}}_2^{(k-B) \times N}
		\end{array}
	\end{array}
\right]
\end{align}
with
\begin{enumerate}
\item $^{*}$\textbf{P}$_{j}^{(N-l) \times N}$ the $(N-l) \times N$ submatrix of \textbf{P}$_{j}$ consisting of its last $N-l$ rows.
\item $\textrm{\textbf{W}}_1^{(k-B) \times (k-B-l)}$ the $(k-B) \times (k-B-l)$ submatrix of \textbf{W} from \eqref{matrix:h-rate} consisting of its $[\delta + (j+1) \cdot N - (k-B+N-l), \delta + j \cdot N -1]$ columns.
\item $\textrm{\textbf{W}}_2^{(k-B) \times N}$ the $(k-B) \times N$ submatrix of \textbf{W} from \eqref{matrix:h-rate} consisting of its $[\delta + j \cdot N, \delta + (j+1) \cdot N -1]$ columns.
\end{enumerate}
Note that  in~\eqref{matrix:G'j-B2}, $\bI_{N-l}$ matrix corresponds to the last $N-l$ symbols $\tilde{\bs}_j$ of the subblock $j$, i.e., $\bs_j = \{s_{\del + j\cdot N},\ldots, s_{\del + (j+1)\cdot N-1}\}$ surely be erased by the burst. We can assume (as worst case) that all symbols associated with $\bI_{k-B}$, i.e., $\bs_L = \{s_B,\ldots, s_{k-1}\}$ are erased. The parity part in~\eqref{matrix:G'j-B2} is not erased  as it happens after the burst. We thus need to argue that one can recover $\tilde{\bs}_j$ and $\bs_L$ upon receiving
\begin{align}
\label{eq:bx_eqn_hr_p2}
\tilde{\bx} = \tilde{\bs}_j \cdot ^{*}\hat{\textbf{P}}_{j}^{(N-l) \times N} + \bs_L \cdot ^{*}\hat{\bW}^{(k-B) \times (k-B+N-l) }
\end{align}
where we have defined
\begin{align}
&^{*}\hat{\textbf{P}}_{j}^{(N-l) \times N}  = 
\begin{bmatrix}
\bm{0}^{(N-l)\times (k-B-l)} &
^{*}\textrm{\textbf{P}}_{j}^{(N-l) \times N}
\end{bmatrix},
&
^{*}\hat{\bW}^{(k-B) \times (k-B+N-l) } =
\begin{bmatrix}
\textrm{\textbf{W}}_1^{(k-B) \times (k-B-l)}	
& \textrm{\textbf{W}}_2^{(k-B) \times N}
\end{bmatrix}.
\end{align}

Now recall that the matrix
\begin{align}
\bG_\mrm{MRD} = 
\left[ 
\begin{array}{c;{2pt/2pt}c}
\begin{array}{c}
\bI_{k-B+\delta}
\end{array}&
\begin{array}{c}
\bP_A^{\del \times B} \\\hdashline[2pt/2pt]
\bW^{(k-B) \times B}\end{array}
\end{array}
\right] =
\left[ 
\begin{array}{c;{2pt/2pt}c}
\begin{array}{c}
\bI_{k-B+\delta}
\end{array}&
\begin{array}{c;{2pt/2pt} c;{2pt/2pt} c}
\bP_A^- & ^{*}\hat{\bP}_A^{\del \times (k-B+N-l)} & \bP_A^+ \\\hdashline[2pt/2pt]
\bW^- & ^{*}\hat{\textbf{W}}^{(k-B) \times (k-B+N-l)}& \bW^+
\end{array}
\end{array}
\right]
\end{align}
is the generator matrix of a Gabidulin code.
Here $ ^{*}\hat{\bP}_A^{\del \times (k-B+N-l)} $ is a submatrix of $\bP_A$ whose columns are aligned with  $ ^{*}\hat{\textbf{W}}^{(k-B) \times (k-B+N-l)}$.
It thus follows that
\begin{align}
\bG_\mrm{MRD}'  =\left[ 
\begin{array}{c;{2pt/2pt} c}
\begin{array}{c}
\bI_{\del}
\\\hdashline[2pt/2pt]
{\bf 0}^{(k-B) \times \del}
\end{array}
&
\begin{array}{c}
^{*}\hat{\bP}_A^{\del \times (k-B+N-l)} \\\hdashline[2pt/2pt]
^{*}\hat{\textbf{W}}^{(k-B) \times (k-B+N-l)}\end{array}
\end{array}
\right] \in \F_{q^m}^{(k-B+\del) \times (k-B+ N-l +\del)}
\end{align}
is the generator matrix of a $(k-B+N-l + \del, k-B+\del)$ MRD code by Lemma~\ref{lem:syscol}.

Since the symbols $\bs_\del$ have already been recovered, the receiver can compute (via~\eqref{eq:bx_eqn_hr_p2})
\begin{align}
\hat{\bx} &= \begin{bmatrix} \bs_\del & \bs_\del \cdot \hat{\bP}_A^{\del \times (k-B+N-l)} + \tilde{\bx}\end{bmatrix} \\
&=  \tilde{\bs}_j \cdot ^{*}\hat{\textbf{P}'}_{j}^{(N-l) \times (k-B+N-l+\del)} + \begin{bmatrix} \bs_\del & \bs_L \end{bmatrix}\cdot  \bG'_\mrm{MRD},
\label{eq:x_expanded_hr3}
\end{align}
where we select
\begin{align}
^{*}\hat{\textbf{P}'}_{j}^{(N-l) \times (k-B+N-l+\del)} = \begin{bmatrix} {\bf 0}^{(N-l) \times \del} &  ^{*}\hat{\textbf{P}}_{j}^{(N-l) \times (k-B+N-l)}\end{bmatrix}.
\end{align}
Since the rank of $^{*}\hat{\textbf{P}'}_{j}^{(N-l) \times (k-B+N-l+\del)} $ is no more than $N-l$ we are guaranteed that there exists a full rank matrix $\bM \in \F_q^{(k-B+N-l+\del) \times (k-B+\del)}$
such that
\begin{align}
^{*}\hat{\textbf{P}'}_{j}^{(N-l) \times (k-B+N-l+\del)}\cdot \bM^{(k-B+N-l+\del) \times (k-B+\del)} = {\bf 0}
\label{eq:cancel_M_hr3}
\end{align}
Multiplying both sides of~\eqref{eq:x_expanded_hr3} by $\bM$ on the right hand side and using~\eqref{eq:cancel_M_hr3} we have that
\begin{align}
\hat{\bx} \cdot \bM = \begin{bmatrix} \bs_\del & \bs_L \end{bmatrix} \cdot \bG'_\mrm{MRD} \cdot \bM.
\end{align}

Finally since $\bG'_\mrm{MRD}$ is a generator matrix of a  MRD code  and $\bM$ is a full rank matrix in the base field the product $\bG'_\mrm{MRD} \cdot \bM$
is invertible by Theorem 1 and thus $\bs_L$ can be uniquely recovered. 

Upon recovery of $\bs_L$ we can cancel its contribution from the right hand side of \eqref{eq:bx_eqn_hr_p2}. It remains to argue that we can uniquely recover 
$\tilde{\bs}_j$ from $\tilde{\bs}_j \cdot ^{*}\hat{\textbf{P}'}_{j}^{(N-l) \times (k-B+N-l+\del)} $. This again follows since any $(N-l) \times (N-l)$ submatrix of 
$^{*}\hat{\textbf{P}'}_{j}^{(N-l) \times (k-B+N-l+\del)}$ is invertible. Indeed it is a square submatrix of $\bP_j$ which is a Cauchy matrix. This establishes the recovery of all erased symbols in $\bs_j$ for $j \in \{1,2,\dots,M-1\}$.

For the other sub-case when we consider a burst beginning in the interval $[\delta,\delta+N-1]$ corresponding to symbols $\bs_0$, we can consider for this setting the following matrix $\bG'_0$ which is also a submatrix of \eqref{matrix:G'-burst}. 

\begin{align}
\label{matrix:G'0-B2}
\textbf{G}'_0 =
\left[
\renewcommand{\arraystretch}{1.5}
	\begin{array}{c;{2pt/2pt}c}
			\begin{array}{c;{2pt/2pt}c}
				\bI_{N-l} & \textbf{0}^{(N-l) \times (\delta + l)}
			\\ 
			\hdashline[2pt/2pt]
				 \textbf{0}^{(\delta + l) \times (N-l)} & \bI_{\small{\delta+l}}		\end{array}&
		\begin{array}{c;{2pt/2pt}c}
			\bm{0}^{(N-l)\times \delta} 
			&
			^{*}\textrm{\textbf{P}}_{0}^{(N-l) \times N}
			\\ 
			\hdashline[2pt/2pt]		
			\textrm{\textbf{W}}_1^{(\delta + l) \times \delta}	
			&
			\textrm{\textbf{W}}_2^{(\delta + l) \times N}
		\end{array}
	\end{array}
\right]
\end{align}
with
\begin{enumerate}
\item $^{*}$\textbf{P}$_{0}^{(N-l) \times N}$ the $(N-l) \times N$ submatrix of \textbf{P}$_{0}$ consisting of its last $N-l$ rows.
\item $\textrm{\textbf{W}}_1^{(\delta + l) \times \delta}$ the $(\delta + l) \times \delta$ submatrix of \textbf{W} from \eqref{matrix:h-rate} consisting of its first $\delta$ columns and last $\delta + l$ rows.
\item $\textrm{\textbf{W}}_2^{(\delta + l) \times N}$ the $(\delta + l) \times N$ submatrix of \textbf{W} from \eqref{matrix:h-rate} consisting of its $[\delta, \delta + N -1]$ columns and last $\delta + l$ rows.
\end{enumerate}
As in~\eqref{matrix:G'j-B2}, $\bI_{N-l}$ matrix corresponds to the last $N-l$ symbols $\tilde{\bs}_0$ of the subblock $0$, i.e., $\bs_0 = \{s_{\del},\dots,  s_{\del + N -1}\}$ which are surely erased by the burst. The difference with  the previous sub-case consists in the fact that among the source symbols $\{s_{B},s_{B+1}\dots,s_{k-1}\}$ only the first $\delta + l$ are erased. Indeed, if we consider the index corresponding to the end of the burst
	$\del + l + B - 1$, only the first $k - 1 - (\del + l + B - 1) = k - B - l - \del$ symbols are erased.
	
By the same argument presented previously in the subsection, we can argue that all symbols in $\bs_0$ are recovered by the deadline of the first symbol in the considered interval.

This conclude the proof of the recovery of all the symbols in the interval $[\delta, B-1]$.	

\end{IEEEproof}

\begin{Lemma}
	Given any burst erasure of length up to $B$, all the source symbols in the interval $[B,k-1]$ are recovered by time $t = B + T - N$ thus before the deadline $t = B + T$ of the first source symbol in the interval of interest.
\end{Lemma}

\begin{IEEEproof}
	All the previous source symbols $s_0, s_1, \dots, s_{B-1}$ are guaranteed to be recovered by time $t = T + B - N$ by Lemma 6 and 7. This time corresponds to the end of the transmitted block and is inferior to the time deadline of the first source symbol given that $N$ is non-negative. We can hence consider for this case the following submatrix of \eqref{matrix:h-rate}
\begin{align}
\label{matrix:G'-final}
\bG' = 
\left[ 
\begin{array}{c;{2pt/2pt}c}
\bI_{k-B} 
&
\bW^{(k-B) \times B}
\end{array}	
\right]
\end{align}

Recall that $\bW^{(k-B) \times B}$ is a part of the parity in the systematic  generator matrix of a Gabidulin code as specified in \eqref{eq:Gmrd_hr} and \eqref{eq:Wselect_HR}. In addition given that Gabidulin codes are also MDS, the parity part of \eqref{eq:Gmrd_hr} $\bP =\left[\begin{array}{c}\bP_A \\\hdashline[2pt/2pt]\bP_B\end{array}\right]$ is a Cauchy matrix. This implies that any square submatrix of $\bP$ is invertible and also by extension any square submatrix of $\bW^{(k-B) \times B}$ is invertible. Hence any pattern of $B$ erasures and especially a burst of size $B$ affecting \eqref{matrix:G'-final} can be fully corrected.
	 
\end{IEEEproof}

\begin{remark}
 Lemma 6,7 and 8 attest that each source	symbol is guaranteed to be recovered by its respective deadline for any burst of maximum length $B$. The recovery of the source symbols in the intervals $[0, \delta-1]$, $[\delta+j \cdot N,\delta +(j+1)\cdot N - 1]$ with $j \in \{0,1,\dots,M-1\}$ and $[B,k-1]$ is achieved by the respective deadline of the first symbol in each of the intervals.
\end{remark}
\section{Comparison of the State of the Art Optimal Streaming Codes}
\subsection{Field Size Requirements}

Recall that in \eqref{matrix:h-rate}, we use a $(2N,N)$ MDS code which requires a field size of $q \geq 2N$ with $q$ being a prime number. In addition, in our construction we also need a $(k-B+\delta,k+\delta)$ Gabidulin code. A generator matrix for such code exists as long as the size of the extension field satisfy $m \geq k+\delta$. Thus our construction requires a field size of
\begin{align}
\begin{split}
	\mathcal{O}\left((2N)^{k+\delta}\right)\leq \mathcal{O}\left((2N)^{T}\right)
\end{split}
\end{align} 
where the inequality comes from the fact that $k \triangleq T - N + 1$ and $\delta \leq N-1$.

\subsection{Comparison with other Constructions}

In this subsection we compare our construction with the two other ones presented in the works of Fong et al. \cite{FKLTZA2018} and Krishnan and Kumar \cite{KK2018}.

Let us reconsider Example 2 with parameters $(W = 11, T = 10, B = 4, N = 2)$ and our explicit matrix construction \textbf{G} given in \eqref{matrix:example2}. 

The reader can observe the difference in structure with the corresponding solution from Fong et al. given in (Lemma 3, \cite{FKLTZA2018}).

\begin{Example}
Possible $\textbf{G}'$ matrix for parameters $(W = 11, T=10, B = 4, N = 2)$ for the construction of Fong et al.
\end{Example}
\begin{align}
\textbf{G}' = 
\left[ 
\begin{array}{c;{2pt/2pt}c}
\begin{array}{c c c c c c c c c}
1 & 0 & 0 & 0 & 0 & 0 & 0 & 0 & 0
\\
0 & 1 & 0 & 0 & 0 & 0 & 0 & 0 & 0
\\
0 & 0 & 1 & 0 & 0 & 0 & 0 & 0 & 0
\\
0 & 0 & 0 & 1 & 0 & 0 & 0 & 0 & 0
\\
0 & 0 & 0 & 0 & 1 & 0 & 0 & 0 & 0
\\
0 & 0 & 0 & 0 & 0 & 1 & 0 & 0 & 0
\\
0 & 0 & 0 & 0 & 0 & 0 & 1 & 0 & 0
\\
0 & 0 & 0 & 0 & 0 & 0 & 0 & 1 & 0
\\
0 & 0 & 0 & 0 & 0 & 0 & 0 & 0 & 1
\end{array}
&
\begin{array}{c}
\begin{array}{c c c c}
\beta_0 & \beta_1 & 0 & 0
\\
0 & \beta_2 & \beta_3 & 0
\end{array}
\\
\hdashline[2pt/2pt]
\begin{array}{c c;{2pt/2pt} c c}
0 & 0 & \beta_4 & \beta_5
\\
0 & 0 & \beta_6 & \beta_7
\end{array}
\\
\hdashline[2pt/2pt]
\begin{array}{c c c c}
1 & 5 & 5^2 & 5^3
\\
1 & 4 & 4^2 & 4^3
\\
1 & 3 & 3^2 & 3^3
\\
1 & 2 & 2^2 & 2^3
\\
1 & 1 & 1 & 1
\end{array}
\end{array}
\end{array}
\right]
\end{align}

The above construction  differs from~\eqref{matrix:example2} as the symbols $\beta_0, \ldots, \beta_7$ are not in a block matrix form. Furthermore \cite{FKLTZA2018} only provides an existential proof of $\beta_0,\ldots, \beta_3$. In contrast all the variables in~\eqref{matrix:example2}  are explicitly determined through the construction of the constituent MDS and MRD codes. On the other hand the field size $q > 2 \left( \binom{T+1}{N} + T - B + 2 \right)$ (Lemma 3, \cite{FKLTZA2018}) required by the construction of Fong et al. is smaller than our exponential $q=\mathcal{O}\left((2N)^{T}\right)$.

Krishnan and Kumar presented in \cite{KK2018} an explicit construction of  optimal codes using a linear combination of MDS code coefficients over an extension field or linearized polynomials evaluations, depending on the given set of parameters $(W,T,B,N)$. Their construction is based on a concatenation approach and is rather different than our technique.

While our proposed construction is restricted to the case when $R \ge 1/2$, we believe this is of practical importance as it uses off-the-shelf MDS and MRD codes as constituent codes in contrast to previous works.


\section{Conclusion}

We propose a new class of streaming codes  for the sliding window erasure channel model $\cC(W,B,N)$. Our work focuses on the case when the code rate is at-least $1/2$. Our construction yields a systematic generator matrix, involves uses off-the-shelf on MDS and Gabidulin codes and achieves optimal error correction. Our proof involves a careful application of the properties of MRD and MDS codes to show successful recovery of the source symbols from all erasure patterns of interest.

As a future work, it will be of interest to examine the case when  $R < \frac{1}{2}$. Indeed such a setting was examined in our investigation. However in our attempts thus far it became necessary to use multiple MRD codes, where the field size of one code is required to be the base field in the subsequent code. This requires a very high field size and thus is not practical. It will be interesting to determine if other approaches involving off-the-shelf MRD can MDS codes could be used in this regime.

\section*{Acknowledgment}
Alessandro Neri pointed the authors to references \cite{Neri2018} and \cite{aless2018systematic}.
 
\newpage
\ifCLASSOPTIONcaptionsoff



%
\fi
\bibliographystyle{IEEEtran}
\bibliography{database}

\begin{thebibliography}{10}
\providecommand{\url}[1]{#1}
\csname url@samestyle\endcsname
\providecommand{\newblock}{\relax}
\providecommand{\bibinfo}[2]{#2}
\providecommand{\BIBentrySTDinterwordspacing}{\spaceskip=0pt\relax}
\providecommand{\BIBentryALTinterwordstretchfactor}{4}
\providecommand{\BIBentryALTinterwordspacing}{\spaceskip=\fontdimen2\font plus
\BIBentryALTinterwordstretchfactor\fontdimen3\font minus
  \fontdimen4\font\relax}
\providecommand{\BIBforeignlanguage}[2]{{%
\expandafter\ifx\csname l@#1\endcsname\relax
\typeout{** WARNING: IEEEtran.bst: No hyphenation pattern has been}%
\typeout{** loaded for the language `#1'. Using the pattern for}%
\typeout{** the default language instead.}%
\else
\language=\csname l@#1\endcsname
\fi
#2}}
\providecommand{\BIBdecl}{\relax}
\BIBdecl

\bibitem{BKTA2013}
A.~Badr, A.~Khisti, W.-T. Tan, and J.~Apostolopoulos, ``Streaming codes for
  channels with burst and isolated erasures,'' in \emph{IEEE INFOCOM}, Turin,
  Italy, Apr. 2013.

\bibitem{FKLTZA2018}
S.~L. Fong, A.~Khisti, B.~Li, W.-T. Tan, X.~Zhu, and J.~Apostolopoulos,
  ``Optimal streaming codes for channels with burst and arbitrary erasures,''
  \emph{accepted to IEEE Trans. on Inf. Theory}, 2018, {\tt arXiv:1801.04241
  [cs.IT]}.

\bibitem{KK2018}
M.~N. Krishnan and P.~V. Kumar, ``Rate-optimal streaming codes for channels
  with burst and isolated erasures,'' 2018, {\tt arXiv:1801.05919 [cs.IT]}.

\bibitem{MartinianSundberg2004}
E.~Martinian and C.-E.~W. Sundberg, ``Burst erasure correction codes with low
  decoding delay,'' vol.~50, no.~10, pp. 2494 -- 2502, 2004.

\bibitem{onewayTransTime}
{International Telecommunication Union}, ``Recommendation {G.114},'' Tech.
  Rep., May 2003.

\bibitem{StockhammerHannuksela2005}
T.~Stockhammer and M.~Hannuksela, ``{H.264/AVC} video for wireless
  transmission,'' vol.~12, pp. 6--13, Aug. 2005.

\bibitem{LQH2013}
D.~Leong, A.~Qureshi, and T.~Ho, ``On coding for real-time streaming under
  packet erasures,'' in \emph{Proc. IEEE Intl. Symp. Inf.~Theory}, Istanbul,
  Turkey, Jul. 2013.

\bibitem{AdlerCassuto2017}
N.~Adler and Y.~Cassuto, ``Burst-erasure correcting codes with optimal average
  delay,'' \emph{IEEE Trans. on Inf. Theory}, vol.~63, no.~5, pp. 2848--2865,
  2017.

\bibitem{rashmi}
M.~Rudow and K.~V. Rashmi, ``Streaming codes for variable-size arrivals,'' in
  \emph{56th Annual Allerton Conference on Communication, Control, and
  Computing, Allerton 2018, Monticello, IL, USA, October 2-5, 2018}, 2018, pp.
  733--740.

\bibitem{KuijperB16}
M.~Kuijper and M.~Bossert, ``On (partial) unit memory codes based on
  reed-solomon codes for streaming,'' in \emph{{IEEE} International Symposium
  on Information Theory, {ISIT} 2016, Barcelona, Spain, July 10-15, 2016},
  2016, pp. 920--924.

\bibitem{napp2016constructing}
D.~Napp and R.~Smarandache, ``Constructing strongly-mds convolutional codes
  with maximum distance profile.'' \emph{Advances in Mathematics of
  Communications}, vol.~10, no.~2, 2016.

\bibitem{almeida2016superregular}
P.~Almeida, D.~Napp, and R.~Pinto, ``Superregular matrices and applications to
  convolutional codes,'' \emph{Linear Algebra and its Applications}, vol. 499,
  pp. 1--25, 2016.

\bibitem{BPKTA17}
A.~Badr, P.~Patil, A.~Khisti, W.-T. Tan, and J.~Apostolopoulos, ``I
  constructions for low-delay streaming codes,'' vol.~63, no.~1, pp. 111 --
  141, 2017.

\bibitem{MacWilliamsSloane1988}
F.~J. MacWilliams and N.~J.~A. Sloane, \emph{The Theory of Error-Correcting
  Codes}, 1st~ed.\hskip 1em plus 0.5em minus 0.4em\relax Amsterdam, Holland:
  North-Holland, Netherlands, 1988.

\bibitem{Gab85}
E.~M. Gabidulin, ``{Theory of codes with maximum rank distance},''
  \emph{Probl.\ Inf.\ Transm.}, vol.~21, no.~1, pp. 1--12, 1985.

\bibitem{Neri2018}
A.~Neri, A.-L. Horlemann-Trautmann, T.~Randrianarisoa, and J.~Rosenthal, ``On
  the genericity of maximum rank distance and gabidulin codes,'' \emph{Designs,
  Codes and Cryptography}, vol.~86, no.~2, pp. 341--363, 2018.

\bibitem{berger2009construction}
T.~P. Berger and A.~Ourivski, ``Construction of new mds codes from gabidulin
  codes,'' in \emph{Proceedings of ACCT}, 2009, pp. 40--47.

\bibitem{Al2017}
A.-L. Horlemann-Trautmann and K.~Marshall, ``New criteria for mrd and gabidulin
  codes and some rank-metric code constructions,'' \emph{Advances in
  Mathematics of Communications}, vol.~11, p. 533, 2017.

\bibitem{aless2018systematic}
A.~Neri, ``Systematic encoders for generalized gabidulin codes and the
  $q$-analogue of cauchy matrices,'' 2018.

\bibitem{Forney1971}
G.~D. Forney, ``Burst-correcting codes for the classic bursty channel,''
  vol.~19, no.~5, pp. 772 -- 781, 1971.

\bibitem{Sil08}
D.~Silva, F.~R. Kschischang, and R.~K\"{o}tter, ``{A rank-metric approach to
  error control in random network coding},'' vol.~54, no.~9, 2008.

\end{thebibliography}

%





\end{document}